\documentclass[preprint,pra]{revtex4}
\usepackage{makeidx}
\usepackage{amssymb}
\usepackage{amsmath}
\usepackage{eurosym}
\usepackage{graphicx}
\usepackage{dcolumn}
\usepackage{bm}

\begin{document}

\title{Solitons and Nonlinear Dynamics in Dual-Core Optical Fibers}
\author{Boris A. Malomed}
\address{Department of Physical Electronics, School of Electrical
Engineering, Faculty of Engineering, Tel Aviv University, Tel Aviv
69978, Israel\\
Laboratory of Nonlinear-Optical Informatics, ITMO University,
St. Petersburg 197101, Russia}

\begin{abstract}
The article provides a survey of (chiefly, theoretical) results obtained for
self-trapped modes (solitons) in various models of one-dimensional optical
waveguides based on a pair of parallel guiding cores, which combine the
linear inter-core coupling with the intrinsic cubic (Kerr) nonlinearity,
anomalous group-velocity dispersion, and, possibly, intrinsic loss and gain
in each core.\ The survey is focused on three main topics: spontaneous
breaking of the inter-core symmetry and the formation of asymmetric temporal
solitons in dual-core fibers; stabilization of dissipative temporal solitons
(essentially, in the model of a fiber laser) by a lossy core
parallel-coupled to the main one, which carries the linear gain; and
stability conditions for $\mathcal{PT}$ (parity-time)-symmetric solitons in
the dual-core nonlinear dispersive coupler with mutually balanced linear
gain and loss applied to the two cores.
\end{abstract}

\maketitle

\textbf{The list of acronyms}

1D: one-dimensional

2D: two-dimensional

BG: Bragg grating

CGLE: complex Ginzburg-Landau equation

CQ: cubic-quintic (nonlinearity)

CW: continuous-wave (solution)

GPE: Gross-Pitaevskii equation

GS: gap soliton

GVD: group-velocity dispersion

MI: modulational instability

NLSE: nonlinear Schr\"{o}dinger equation

$\mathcal{PT}$: parity-time (symmetry)

SBB: symmetry-breaking bifurcation

SP: solitary pulse

SPM: self-phase modulation

VA: variational approximation

WDM: wavelength-division multiplexing

XPM: cross-phase modulation

\section{Introduction}

One of basic types of optical waveguides is represented by dual-core
couplers, in which parallel guiding cores exchange the propagating
electromagnetic fields via evanescent fields tunneling across the dielectric
barrier separating the cores \cite{Huang}. In most cases, the couplers are
realized as twin-core optical fibers \cite{fiber-coupler,fiber-coupler2},
or, in a more sophisticated form, as twin-core structures embedded in
photonic-crystal fibers \cite{PCF}. Such double fibers can be drawn by means
of an appropriately shaped preform from melt, or fabricated by pressing
together two single-mode fibers, with the claddings removed in the contact
area. Alternatively, a microstructured fiber with a dual guiding core can be
fabricated and used too \cite{Russell}.

If the intrinsic nonlinearity in the cores is strong enough, the power
exchange between them is affected by the intensity of the guided signals
\cite{Jensen}. This effect may be used as a basis for the design of diverse
all-optical switching devices \cite{switch1}-\cite{review} and other
applications, such as nonlinear amplifiers \cite{amplifier,amplifier2},
stabilization of wavelength-division-multiplexed (WDM) transmission schemes
\cite{WDM}, logic gates \cite{logic}, and bistable transmission \cite{Leon}.
Nonlinear couplers also offer a setup for efficient compression of solitons
by passing them into a fiber with a smaller value of the group-velocity
dispersion (GVD) coefficient: as demonstrated in work \cite{HH}, the highest
quality of the soliton's compression is achieved when two fibers with
different dispersion coefficients are not directly spliced one into the
other, but are connected so as to form a coupler (a necessarily asymmetric
one, in this case).

In addition to the simplest dual-core system, realizations of nonlinear
couplers have been proposed in many other settings, including the use of the
bimodal structure (orthogonal polarizations) of guided light \cite{Trillo},
semiconductor waveguides \cite{semi}, plasmonic media \cite%
{plasma1,plasma2,plasma3}, and twin-core Bragg gratings \cite%
{Bragg,Tsofe,Sukhorukov}, to mention just a few. In addition to the
ubiquitous Kerr (local cubic) nonlinearity of the core material, the
analysis has been developed for systems with nonlinearities of other types,
including saturable \cite{satur}, quadratic (alias
second-harmonic-generating) \cite{chi2,Shapira}, cubic-quintic (CQ) \cite%
{Albuch}, and nonlocal cubic interactions \cite{Nonlocal}. Unlike the Kerr
nonlinearity, more general types of the self-interaction of light can be
realized not in fibers (i.e., not in the \textit{temporal domain}), but
rather in planar waveguides (i.e., in the \textit{spatial domain}).
Theoretical modeling of these settings is facilitated by the fact that the
respective nonlinear Schr\"{o}dinger equations (NLSEs)\ for the evolution of
the local amplitude of the electromagnetic waves takes identical forms in
the temporal and spatial domains, with the temporal variable replaced by the
transverse spatial coordinate, in the latter case. Generally, dual-core
systems are adequately modeled by systems of two linearly coupled NLSEs, in
which the linear coupling represents the tunneling of electromagnetic fields
between the cores \cite{Jensen,Wabnitz,Snyder}.

Further, effective \emph{discretization} of continuous nonlinear couplers
may be provided, in the spatial domain too, by the consideration of parallel
arrays of discrete waveguides \cite{Herring}-\cite{Fangwei}. The coupler
concept was also extended for the \textit{spatiotemporal }propagation of
light in dual-core planar waveguides, with the one-dimensional (1D) NLSE
replaced by its two-dimensional (2D) version, which includes both the
temporal and spatial transverse coordinates \cite{Dror}.

Similar to the double-fiber waveguides for optical waves are dual-core
cigar-shaped (strongly elongated) traps for matter waves in atomic
Bose-Einstein condensates (BECs) \cite{Randy}. Transmission of matter waves
in these settings have been studied theoretically \cite{BEC1,Luca,Marek},
making use of the fact that the Gross-Pitaevskii equation (GPE)\ for the
mean-field wave function of the matter waves in BEC \cite{GPE} is actually
identical to the NLSE for electromagnetic waves in similar optical
waveguides.

The above-mentioned systems in optics and BEC imply lossless propagation of
optical and atomic waves, hence the respective models are based on the NLSEs
and GPEs which do not include dissipative terms. On the other hand,\ loss
and gain play an important role in many optical systems, such as fiber
lasers. The fundamental model of such systems is based on complex
Ginzburg-Landau equations (CGLEs), i.e., an extension of the NLSE with real
coefficients replaced by their complex counterparts \cite{CGL,CGL2}.
Accordingly, dual-core fiber lasers are described by systems of linearly
coupled CGLEs \cite{Sigler}. Dissipative linearly coupled systems with the
gain and loss applied to different cores are relevant too, as models
admitting stable transmission \cite{Winful,Atai1,Atai2}, filtering \cite%
{filter}, and nonlinear amplification \cite{nonlin-ampl} of optical pulses
in fiber lasers with the cubic nonlinearity, see a brief review of the topic
in \cite{Chaos-review}.

A special system is one with exactly equal gain and loss acting in the
parallel-coupled cores, which are identical as concerns other coefficients
\cite{PT1}. Such settings feature the $\mathcal{PT}$ (parity-time) symmetry
between the cores (for the general concept of the $\mathcal{PT}$ symmetry,
see original works \cite{Bender}-\cite{Longhi} and reviews \cite%
{Bender-review}-\cite{review2}). The linear spectrum of the $\mathcal{PT}$%
-symmetric coupler may remain purely real (i.e., it does not produce decay
due to imbalanced loss or blowup due to imbalanced gain), provided that the
gain-loss coefficient does not exceed a critical value (in fact, it is
exactly equal to the coefficient of the linear coupling between the cores
\cite{PT1,PT2}; see Section III below). Overall, $\mathcal{PT}$-symmetric
settings may be considered as dissipative systems which are able to emulate
conservative ones, as they support not only real spectra but also stable
soliton families if appropriate nonlinearity is included \cite%
{Chr2,review1,review2}, as is shown in detail below in Section III (generic
dissipative systems create isolated nonlinear states (in particular,
dissipative solitons \cite{me}), which play the role of attractors, rather
than continuous families of stable solutions).

Thus, nonlinear dual-core systems represent a vast class of settings
relevant to optics, BEC, and other areas, which offer a possibility to model
and predict many physically significant effects. As examples of similar
systems which are well known in completely different areas of physics, it is
relevant to mention tunnel-coupled pairs of long Josephson junction, which
are described by systems of linearly coupled sine-Gordon equations (see
original works \cite{JJ1}-\cite{JJ3} and reviews \cite{JJ-review1,JJ-review2}%
), and the propagation of internal waves in stratified liquids with two
well-separated interfaces, which are described by pairs of linearly-coupled
Korteweg de Vries equations, which were derived in various forms \cite{KdV1}-%
\cite{Mexico}. The purpose of this article is to present a reasonably
compact review of the corresponding models and results. Because the general
topic is very broad, the review is limited to optical waveguides based on
dual-core optical fibers. Related settings, such as those based on double
planar waveguides and double traps for matter waves in BEC, are briefly
mentioned in passing.

A fundamental property of nonlinear couplers with symmetric cores is the
\textit{symmetry-breaking bifurcation} (SBB), which destabilizes obvious
symmetric modes (sometimes called \textit{supermodes}, as they extend to
both individual cores, which support individual modes), and gives rise to
asymmetric states. The SBB was theoretically analyzed in detail for
temporally uniform states (alias \textit{continuous waves}, CWs) in
dual-core nonlinear optical fibers \cite{Snyder}, and, in parallel, for
self-trapped solitary waves, i.e., temporal solitons in the same system \cite%
{Wabnitz}-\cite{Skinner}, as well as for dual-core nonlinear fibers with
Bragg gratings (BGs) written on each core \cite{Bragg}. Some results
obtained in this direction were summarized in early review \cite{Wabnitz2},
and later in \cite{Progress}. The SBB\ analysis was then extended to
solitons in couplers with the quadratic \cite{chi2} and CQ \cite{Albuch}
nonlinearities.

The Kerr nonlinearity in the dual-core system gives rise to the \emph{%
subcritical} SBB for solitons, with originally unstable branches of emerging
asymmetric modes going backward (in the direction of weaker nonlinearity)
and then turning forward \cite{bifurcations}. The asymmetric modes retrieve
their stability at the turning points. On the other hand, the \emph{%
supercritical} SBB gives rise to stable branches of asymmetric solitons
going in the forward direction. For solitons, the SBB of the latter type\
occurs in twin-core Bragg grating \cite{Bragg} (see subsection II.D below),
and in the system with the quadratic nonlinearity \cite{chi2}. The coupler
with the intra-core CQ nonlinearity gives rise to a closed \emph{bifurcation
loop}, whose shape may be concave or convex \cite{Albuch} (see details in
subsection II.E below).

In models of nonlinear dual-core couplers, the SBB point can be found in an
exact analytical form for the system with the cubic nonlinearity \cite%
{Wabnitz}, and the emerging asymmetric modes were studied by means of the
variational approximation (VA) \cite{Laval,Muschall,satur,chi2,Bragg,Dror}
and numerical calculations \cite{Akhm,Akhm2}, see also reviews \cite%
{Wabnitz2} and \cite{Progress}.

In addition to the studies of solitons in uniform dual-core systems, the
analysis was developed for \emph{fused couplers}, in which the two cores are
joined in a narrow segment \cite{X}. In the simplest approximation, the
corresponding dependence of the coupling strength on coordinate $z$ may be
approximated by the delta-function, $\delta (z)$. Originally, interactions
of solitons with a locally fused segment were studied in the temporal
domain, \textit{viz}., for bright \cite{X,Manolo,India} and dark \cite{dark}
solitons in dual-core optical fibers and fiber lasers \cite{laser}. In that
case, the coupling affects the solitons only in the course of a short
interval of their evolution. The \textit{spatial-domain} optical counterpart
of the fused coupler is provided by a dual-core planar waveguide with a
narrow coupling segment created along the coordinate ($x$) perpendicular to
the propagation direction ($z$) \cite{Akhmed,Raymond}. Dynamics of spatial
optical solitons in such settings, including stationary solitons trapped by
the fused segment of the coupler, and scattering of incident solitons on one
or several segments, was analyzed in \cite{Alon}.

The objective of this article is to present\ a review of basic findings
produced by studies of models developed for dual-core nonlinear optical
fibers and fiber lasers, along the above-mentioned directions. The review is
chiefly focused on theoretical results, as experimental ones are still
missing for solitons, in most cases. In Section II, the most fundamental
results are summarized for the SBBs of solitons in dual-core fibers with
identical cores. Some essential findings for asymmetric waveguides, with
different cores, are included too. The results are produced by a combination
of numerical and methods and analytical approximations (primarily, the
variational approximation, VA). In Section III, the results are presented
for the creation of stable dissipative solitons in models of fiber lasers,
the stabilization being provided by coupling the main core, which carries
the linear gain, to a parallel lossy one. Section IV is focused on nonlinear
dual-core couples featuring the above-mentioned parity-time ($\mathcal{PT}$)
symmetry, which is provided by creating mutually balanced gain and loss in
two otherwise identical cores, the main issue being stability conditions for
the corresponding $\mathcal{PT}$-symmetric solitons (which can be solved in
an exact analytical form, in this model). The article is concluded by
Section V.

\section{Solitons in dual-core fibers}

\subsection{The symmetry-breaking bifurcation (SBB)\ of solitons}

\subsubsection{The formulation of the model}

The basic model of the symmetric coupler, i.e., a dual-core fiber with equal
dispersion and nonlinearity coefficients in the parallel-coupled cores, is
represented by a system of linearly coupled NLSEs, which are written here in
the scaled form \cite{fiber-coupler2}, with subscripts standing for partial
derivatives:
\begin{eqnarray}
iu_{z}+\frac{1}{2}u_{\tau \tau }+|u|^{2}u+Kv &=&0,  \label{ucoupler} \\
iv_{z}+\frac{1}{2}v_{\tau \tau }+|v|^{2}v+Ku &=&0,  \label{vcoupler}
\end{eqnarray}%
where $z$ is the propagation distance, $\tau \equiv t-z/V_{\mathrm{gr}}$ is
the reduced time ($t$ is the physical time, and $V_{\mathrm{gr}}$ is the
group velocity of the carrier wave \cite{Agrawal}), $u$ and $v$ are
amplitudes of the electromagnetic waves in the two cores, sign $+$ in front
of the group-velocity-dispersion (GVD) terms, represented by the second
derivatives, implies the anomalous character of the GVD in the fiber \cite%
{Agrawal}, the cubic terms represent the intra-core Kerr effect, and $K$,
which is defined to be positive (actually, it may be scaled to $K\equiv 1$)
is the coupling constant accounting for the light exchange between the
cores.\

An additional effect which can be included in the model represents the
temporal dispersion of the inter-core coupling, represented by its own real
coefficient, $K^{\prime }$. The accordingly modified Eqs. (\ref{ucoupler})
and (\ref{vcoupler}) take the form \cite{Kin}%
\begin{eqnarray}
iu_{z}+\frac{1}{2}u_{\tau \tau }+|u|^{2}u+Kv+iK^{\prime }v_{\tau } &=&0,
\label{Kin-u} \\
iv_{z}+\frac{1}{2}v_{\tau \tau }+|v|^{2}v+iKu+K^{\prime }u_{\tau } &=&0.
\label{Kin-v}
\end{eqnarray}%
Below, this generalization of the nonlinear-coupler model is not considered
in detail, as detailed analysis has demonstrated that the dispersion of the
inter-core coupling does not produce drastic changes in properties of
solitons \cite{Kin-Akhm}.

Equations (\ref{ucoupler}) and (\ref{vcoupler}) can be derived, by means of
the standard variational procedure, from the respective Lagrangian ($L$),
which, in turns, includes the Hamiltonian of the system ($H$) \cite{Progress}%
:
\begin{eqnarray}
L &=&\int_{-\infty }^{+\infty }\left[ \frac{i}{2}\left( u^{\ast
}u_{z}+v^{\ast }v_{z}\right) d\tau +\mathrm{c.c.}\right] -H,  \label{L} \\
H &=&\int_{-\infty }^{+\infty }\left[ \frac{1}{2}\left( \left\vert u_{\tau
}\right\vert ^{2}+\left\vert v_{\tau }\right\vert ^{2}\right) -\frac{1}{2}%
\left( |u|^{4}+|v|^{4}\right) -K\left( u^{\ast }v+uv^{\ast }\right) \right] ,
\label{H}
\end{eqnarray}%
where both $\ast $ and c.c. stand for the complex-conjugate expressions. The
Hamiltonian, along with the integral energy (alias total norm) of the
solution,%
\begin{equation}
E=\frac{1}{2}\int_{-\infty }^{+\infty }\left( |u|^{2}+|v|^{2}\right) d\tau ,
\label{E}
\end{equation}%
and the total momentum,%
\begin{equation}
P=\frac{i}{2}\int_{-\infty }^{+\infty }\left[ \left( uu_{\tau }^{\ast
}+vv_{\tau }^{\ast }\right) +\mathrm{c.c.}\right] d\tau ,  \label{P}
\end{equation}%
are dynamical invariants (conserved quantities) of the system.

If the dispersion of the inter-core dispersion is included, see Eqs. (\ref%
{Kin-u}) and (\ref{Kin-v}), the additional term in the Hamiltonian density
in Eq. (\ref{H}) is $-\left( iK/2\right) \left( u^{\ast }v_{\tau }+v^{\ast
}u_{\tau }-uv_{\tau }^{\ast }-vu_{\tau }^{\ast }\right) $, while the
expression for the total energy and momentum keep\ the same form as defined
in Eqs. (\ref{E}) and (\ref{P}).

Equations (\ref{ucoupler}) and (\ref{vcoupler}) admit obvious symmetric and
antisymmetric soliton solutions (\emph{supermodes}),%
\begin{equation}
u=\pm v=a^{-1}\mathrm{sech}\left( \frac{\tau }{a}\right) \exp \left( \frac{iz%
}{2a^{2}}\pm iKz\right) ,  \label{+_}
\end{equation}%
where $a$ is an arbitrary width, which determines the total energy (\ref{E})
and Hamiltonian (\ref{H}) of the symmetric and antisymmetric states:%
\begin{equation}
E_{\mathrm{symm-sol}}=2a^{-1},~H_{\mathrm{symm-sol}}=-\frac{2}{3}a^{-3}\mp
4Ka^{-1}.  \label{Esymm}
\end{equation}%
In the case of $K>0$ (that may always be fixed by definition), the
antisymmetric solitons are unstable \cite{Akhm2}, as they correspond to a
maximum, rather than minimum, of the coupling term ($\sim K$) in Hamiltonian
(\ref{Esymm}), therefore they are not considered below. For the symmetric
solitons, the SBB, which destabilizes them and replaces them by stable
asymmetric solitons, with different energies in the two cores, is an issue
of major interest. The onset of the SBB of the symmetric soliton, i.e., the
value of the soliton's energy at the SBB point, can be found in an exact
analytical form \cite{Wabnitz}. To this end, one looks for a general
(possibly asymmetric) stationary solution of Eqs. (\ref{ucoupler}) and (%
\ref{vcoupler}) for solitons with propagation constant $k$\ as%
\begin{equation}
\left\{ u\left( z,\tau \right) ,v\left( z,\tau \right) \right\}
=e^{ikz}\left\{ U(\tau ),V(\tau )\right\} ,  \label{uvUV}
\end{equation}%
where real functions $U$ and $V$ satisfy the following ordinary differential
equations:%
\begin{eqnarray}
\frac{1}{2}\frac{d^{2}U}{d\tau ^{2}}+U^{3}+KV &=&kU,  \label{U} \\
\frac{1}{2}\frac{d^{2}V}{d\tau ^{2}}+V^{3}+KU &=&kV.  \label{V}
\end{eqnarray}

The SBB corresponds to the emergence of an \emph{antisymmetric} eigenmode of
infinitesimal perturbations,%
\begin{equation}
\left\{ \delta U(\tau ),\delta V(\tau )\right\} =\varepsilon \left\{
U_{1}(\tau ),-U_{1}(\tau )\right\}   \label{anti}
\end{equation}%
($\varepsilon $ is a vanishingly small perturbation amplitude) around the
unperturbed symmetric solution of Eqs. (\ref{U}) and (\ref{V}), which is
taken as per Eq. (\ref{+_}), i.e.,
\begin{equation}
U_{0}(\tau )=V_{0}(\tau )\equiv a^{-1}\mathrm{sech}\left( \frac{\tau }{a}%
\right) ,~k=\frac{1}{2a^{2}}+K.  \label{U0}
\end{equation}%
The linearization of Eqs. (\ref{U}) and (\ref{V}) around the exact symmetric
state leads to the equation%
\begin{equation}
\frac{1}{2}\frac{d^{2}U_{1}}{d\tau ^{2}}+3U_{0}^{2}U_{1}-\left( K+k\right)
U_{1}=0,  \label{U1}
\end{equation}%
which is tantamount to the stationary version of the solvable 1D linear
Schr\"{o}dinger equation with the P\"{o}schl-Teller potential. Then, with the
help of well-known results from quantum mechanics \cite{LL}, it is easy to
find that, with the growth of the soliton's energy $E$, i.e., with the
increase of $k$ (see Eqs. (\ref{Esymm}) and (\ref{U0})), a nontrivial
eigenstate, produced by Eq. (\ref{U1}), appears at \cite{Wabnitz}%
\begin{equation}
E=E_{\mathrm{bif}}\equiv 4\sqrt{K/3}\approx 2.31\sqrt{K}.  \label{E2}
\end{equation}%
Thus, the SBB\ and destabilization of the symmetric solitons (\ref{U0}) take
place precisely at point (\ref{E2}).

\subsubsection{Continuous-wave (CW) states and their modulational
instability (MI)}

To complete the formulation of the model of the symmetric nonlinear coupler,
it is relevant to mention that, besides the solitons, it admits simple
continuous-wave (CW) states, with constant $U$ and $V$ in Eq. (\ref{uvUV}).
Indeed, Eqs. (\ref{U}) and (\ref{V}) easily produce the full set of CW
solutions: symmetric and antisymmetric ones,
\begin{equation}
U_{\mathrm{symm}}^{\mathrm{(CW)}}=V_{\mathrm{symm}}^{\mathrm{(CW)}}=\sqrt{k-K%
},~U_{\mathrm{anti}}^{\mathrm{(CW)}}=-V_{\mathrm{anti}}^{\mathrm{(CW)}}=%
\sqrt{k+K},  \label{symm&anti}
\end{equation}%
which exist, respectively, at $k>K$ and $k>-K$, respectively. With the
growth of $k$, i.e., increase of the CW amplitude, the symmetric state
undergoes the SBB at $k=2K$, giving rise to asymmetric CW states, which
exist at $k>2K$:%
\begin{equation}
U_{\mathrm{asymm}}^{\mathrm{(CW)}}=\sqrt{\frac{k}{2}+\sqrt{\frac{k^{2}}{4}%
-K^{2}}},~V_{\mathrm{asymm}}^{\mathrm{(CW)}}=\sqrt{\frac{k}{2}-\sqrt{\frac{%
k^{2}}{4}-K^{2}}}~  \label{asymm}
\end{equation}%
(and its mirror image, with $U\rightleftarrows V$). The SBB\ for CW states
in models of couplers with more general nonlinearities was studied in detail
in Ref. \cite{Snyder}.

However, all the CW states are subject to the \textit{modulational
instability} (MI) \cite{MI-Rome}, which, roughly speaking, tends to split
the CW into a chain of solitons. While this conclusion is not surprising in
the case of the anomalous GVD in Eqs. (\ref{ucoupler}) and (\ref{vcoupler}),
as it gives rise to the commonly known MI in the framework of the single
NLSE \cite{Agrawal}, all symmetric and asymmetric CW states are
modulationally unstable too in the system of linearly coupled NLSEs with the
normal sign of the GVD in each equation \cite{Tasgal}, this MI being
produced by the linear coupling. Because of the instability of the CW
background, nonlinear couplers, even with normal GVD, cannot support stable
dark solitons or domain walls, i.e., delocalized states in the form of two
semi-infinite asymmetric CWs, which are transformed into each other by
substitution $U\rightleftarrows V$, linked by a transient layer \cite{DW}.
Because bright solitons cannot exist in the case of the normal GVD, the
development of the MI in the latter case leads to a state in the form of
\textquotedblleft optical turbulent" \cite{Tasgal}.

As concerns the temporal dispersion of the inter-core coupling (see Eqs. (%
\ref{Kin-u}) and (%
\ref{Kin-v})), its effect on the MI of the CW states was studied in Ref. %
\cite{Kin-MI}.

\subsubsection{The variational approximation (VA) for solitons}

Asymmetric solitons, which emerge at the SBB point, cannot be found in an
exact form, but they can be studied by means of the VA. This approach for
solitons in nonlinear couplers was developed in works \cite{Laval}-\cite%
{UNSW}, see also work \cite{Ank} which discussed limitations of the VA in
this setting. Here, the main findings produced by the VA, and their
comparison with results obtained by means of numerical methods are presented
as per work \cite{Skinner}.

The VA is based on the following trial analytical form (\emph{ansatz}) for
the two-soliton soliton:
\begin{eqnarray}
u &=&A\cos (\theta )\mathrm{sech}\left(
\frac{\tau }{a}\right) \exp \left( i(\phi +\psi )+ib\tau ^{2}\right) ,
\label{uansatzcoupler} \\
v &=&A\sin (\theta )\mathrm{sech}\left( \frac{\tau }{a}\right) \exp \left(
i(\phi -\psi )+ib\tau ^{2}\right) .  \label{vansatzcoupler}
\end{eqnarray}%
where real variational parameters, $A$, $\theta $, $a$, $\phi $, $\psi $, $b$%
, may be functions of the propagation distance, $z$. In particular, $A(z)$
and $a(z)$ are common amplitude and width of the two components, chirp $b(z)$
must be introduced, as it is well known \cite{Anderson,Anderson2}, in the
dynamical ansatz which allows evolution of the soliton's width, $\phi (z)$
is an overall phase of the two-component soliton, angle $\theta (z)$
accounts for the distribution of the energy between the components, and $%
\psi (z)$ is a relative phase between them. The shape and phase parameters
form conjugate pairs, \textit{viz}., $\left( A,\phi \right) $, $\left(
\theta ,\psi \right) $, and $\left( a,b\right) $.

Note that the ansatz based on Eqs. (\ref{uansatzcoupler}) and (\ref%
{vansatzcoupler}) assumes that centers of the two components of the soliton
are stuck together. This implies that the linear coupling between the two
cores is strong, which corresponds to the real physical situation.
Nevertheless, it is also possible to consider a case when the linear
coupling plays the role of a small perturbation, making a two-component
soliton a weakly bound state of two individual NLSE solitons belonging to
the two cores \cite{Abd,KM,Cohen}.

Switching of a soliton between the two cores of the coupler was considered,
on the basis of a full system of variational equations for ansatz (\ref%
{uansatzcoupler}), (\ref{vansatzcoupler}) in work \cite{Uzunov}. It was also
demonstrated in work \cite{Smyth} that the approximation for the switching
dynamics in the nonlinear coupler can be further improved if the radiation
component of the wave field is incorporated into the ansatz.

Here, the consideration is focused on the basic case of static solitons, for
which ansatz (\ref{uansatzcoupler}), (\ref{vansatzcoupler}) gives rise to
the following variational equations, in which all parameters of ansatz (\ref%
{uansatzcoupler})-(\ref{vansatzcoupler}), but overall phase $\phi $, are
assumed constant:
\begin{gather}
\sin (2\theta )\sin (2\psi )=0,  \label{coupler1} \\
\frac{E}{3a}\cos (2\theta )-K\,\mathrm{cot}(2\theta )\cos (2\psi )=0,
\label{coupler2} \\
a^{-1}=E\left[ 1-\frac{1}{2}\sin ^{2}(2\theta )\right] ,  \label{coupler3}
\end{gather}%
\begin{equation*}
\frac{d\phi }{dz}\,=\,-\,\frac{1}{6a^{2}}+\frac{2E}{3a}\left( 1-\frac{1}{2}%
\sin ^{2}(2\theta )\right) +\kappa \sin (2\theta )\cos (2\psi ),
\end{equation*}%
where $E$ is the soliton's energy, which, according to its definition (\ref%
{E}), takes value $E=A^{2}a$ for ansatz (\ref{uansatzcoupler}), (\ref%
{vansatzcoupler}).

As it follows from Eq. (\ref{coupler1}), the static soliton may have either $%
\sin (2\theta )=0$, or $\sin (2\psi )=0$. According to the underlying
ansatz, the former solution implies that all the energy resides in a single
core, which contradicts Eqs. (\ref{ucoupler}) and (\ref{vcoupler}), hence
this solution is spurious. The latter solution, $\sin (2\psi )=0$, implies
that $\cos (2\psi )=\pm 1$. As mentioned above, the solutions corresponding
to $\cos (2\psi )=-1$, i.e., antisymmetric ones, with respect to the two
components, are unstable. Therefore, only the case of $\cos (2\psi )=+1$,
corresponding to solitons with in-phase components, is considered here.
Then, width $a$ can be eliminated by means of Eq. (\ref{coupler3}), and the
remaining equation (\ref{coupler3}) for the energy-distribution angle $%
\theta $ takes the form of
\begin{equation}
\cos (2\theta )\left[ \frac{E^{2}}{3K}\sin (2\theta )\left( 1-\frac{1}{2}%
\sin ^{2}(2\theta )\right) -1\right] =0.  \label{couplertheta}
\end{equation}

Further analysis reveals that, in the interval $0<E^{2}<E_{1}^{2}$, where
\begin{equation}
E_{1}^{2}=(9/4)\sqrt{6}K\approx \allowbreak 5.\,\allowbreak 511\,K,
\label{couplerE1}
\end{equation}%
the only relevant solution to Eq. (\ref{couplertheta}) is the symmetric one,
with $\theta =\pi /4$ (corresponding to $\cos \left( 2\theta \right) =0$)
and equal energies in both components, according to Eqs. (\ref%
{uansatzcoupler}) and (\ref{vansatzcoupler}). When the soliton's energy
attains value $E_{1}$, predicted by Eq. (\ref{couplerE1}), there emerge
\emph{asymmetric} solutions with $\cos (2\theta )=\pm 1/\sqrt{3}$. When $%
E^{2}$ attains a slightly larger value,
\begin{equation}
E_{2}^{2}=6K,  \label{couplerE2}
\end{equation}%
a \emph{backward (subcritical) bifurcation} \cite{bifurcations} occurs,
which makes the symmetric solution with $\theta =\pi /4$ unstable. The
comparison with full numerical results corroborates the weakly subcritical
shape of the SBB for solitons in the nonlinear coupler.

\begin{figure}[tbp]
\begin{center}
\includegraphics[height=10cm]{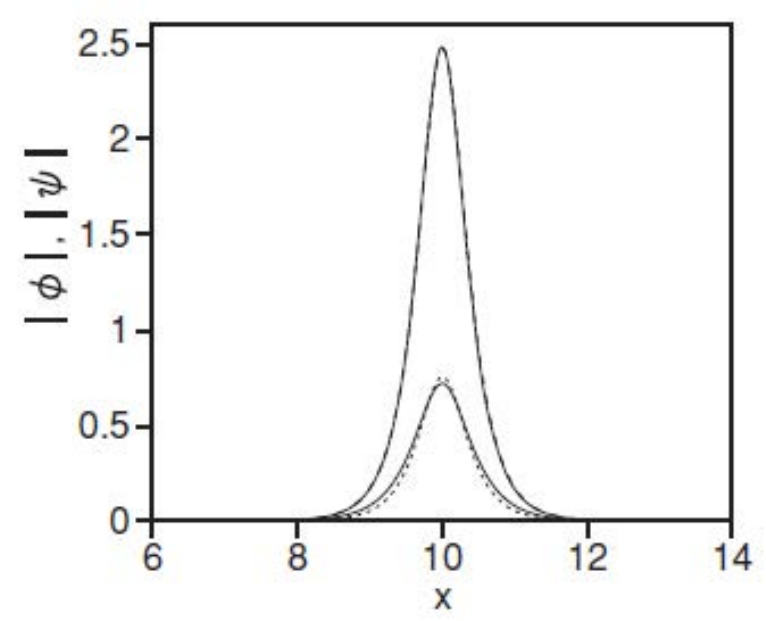}
\end{center}
\caption{A typical example of two components of a stable asymmetric soliton,
with $\left\vert \protect\phi (x)\right\vert \equiv U(\protect\tau )$, $%
\left\vert \protect\psi (x)\right\vert \equiv V(\protect\tau )$, as per Ref.
\protect\cite{HS}. Continuous and dashed lines designate the numerically
found solution and its VA-produced counterpart, respectively.}
\label{fig1}
\end{figure}

\begin{figure}[tbp]
\begin{center}
\includegraphics[height=10cm]{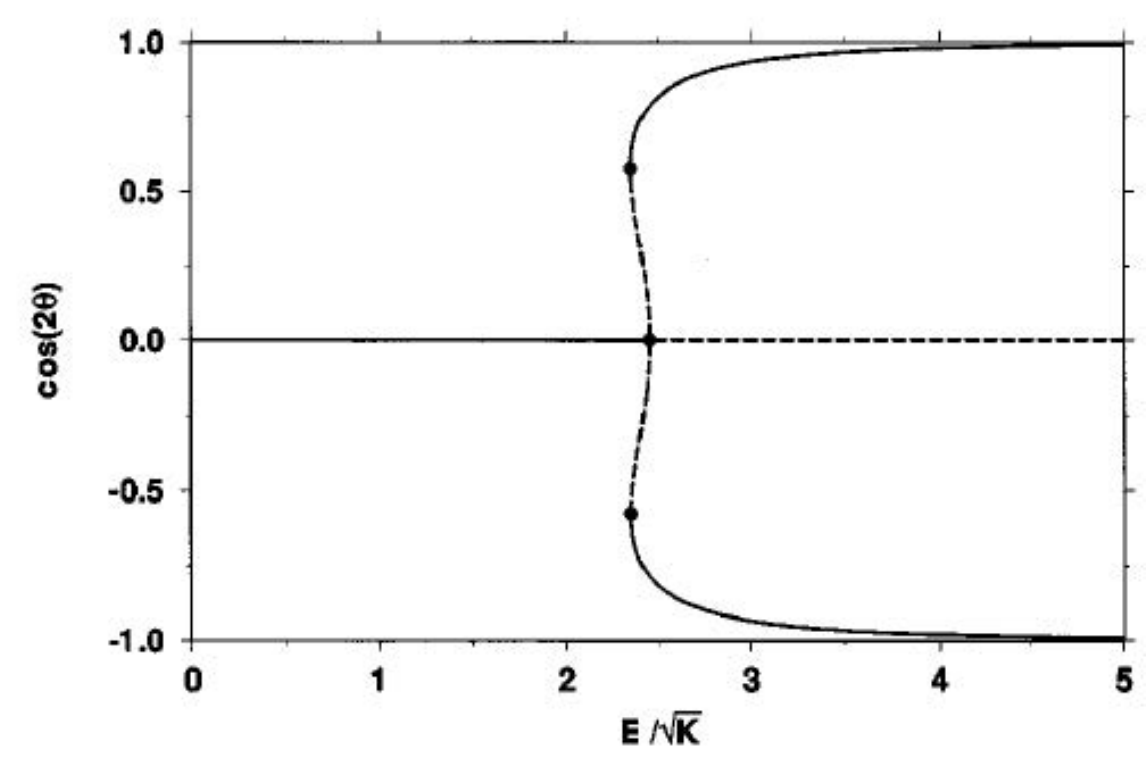}
\end{center}
\caption{The dependence of the asymmetry parameter of two-component solitons
in the nonlinear coupler with identical cores, $\cos (2\protect\theta )$, on
the scaled total energy, $E/\protect\sqrt{K}$, as predicted by the VA, see
Eq. (\protect\ref{couplerasymmetry}). The figure demonstrates a weakly
subcritical SBB, solid and dashed lines designating stable and unstable
states, respectively. The results are presented as per Ref. \protect\cite%
{Skinner}.}
\label{fig2}
\end{figure}

A typical example of an asymmetric soliton is displayed in Fig. \ref{fig1},
and the entire bifurcation diagram is presented in Fig. \ref{fig2}. Note
that quantity $\cos (2\theta )$, which is used as the vertical coordinate in
the diagram, measures the asymmetry of the soliton, because, as it follows
from Eqs. (\ref{uansatzcoupler}) and (\ref{vansatzcoupler}),
\begin{equation}
\cos (2\theta )\equiv \frac{E^{(1)}-E^{(2)}}{E^{(1)}+E^{(2)}},
\label{couplerasymmetry}
\end{equation}%
where $E^{(j)}$ is the energy in the $j$-th core. Even without detailed
stability analysis, one can easily distinguish between stable and unstable
branches in the diagram, using elementary theorems of the bifurcation theory
\cite{bifurcations}.

Thus, the VA predicts the backward bifurcation at the soliton's energy $%
E_{2}=\sqrt{6K}\approx 2.45\sqrt{K}$. The accuracy of the VA is
characterized by comparison of this prediction with the above-mentioned
exactly found bifurcation value (\ref{E2}), the relative error being $0.057$
(the analytical solutions for $E_{\mathrm{bif}}$ does not predict the
subcritical character of the SBB).

The above consideration addressed a single soliton in the nonlinear coupler.
A cruder version of the VA was used in work \cite{Doty1995} to analyze
two-soliton interactions in the same system. Accurate numerical results for
the interaction were reported in \cite{PMC1998}. Furthermore, it was
recently demonstrated that chains of stable solitons with opposite signs
between adjacent ones, in the dual-core fiber, support the propagation of
\emph{supersolitons}, i.e., self-trapped collective excitations in the chain
of solitons \cite{LuLi}. The underlying chain may be built of symmetric
solitons, as well as of asymmetric ones, with alternating polarities, i.e.,
placements of larger and smaller components in the two cores.

\subsection{Gap solitons in asymmetric dual-core fibers}

Asymmetric dual-core fibers, consisting of two different cores, can be
easily fabricated, and properties of solitons in them may be markedly
different from those in the symmetric couplers. A general model of the
asymmetric coupler is (cf. Eqs. (\ref{ucoupler}), (\ref{vcoupler}))
\begin{gather}
iu_{z}+qu+\frac{1}{2}u_{\tau \tau }+|u|^{2}u+v=0,  \label{uasymm} \\
iv_{z}-\delta \cdot \left( qv+\frac{1}{2}v_{\tau \tau }\right) +|v|^{2}v+u=0,
\label{vasymm}
\end{gather}%
where real parameter $-\delta $ accounts for the difference between GVD
coefficients in the cores, and another real coefficient, $(1+\delta )q$,
defines the phase-velocity mismatch between them, while a possible
group-velocity mismatch can be eliminated in the equations by a simple
transformation.

The effect of the asymmetry between the cores on the SBB for solitons was
addressed in Refs. \cite{Skinner} and \cite{Taras} (strictly speaking, in
this case the subject of the analysis is spontaneous breaking of \emph{%
quasi-symmetry}, which remains after lifting the exact symmetry by the
mismatch between the cores). In \cite{Taras}, a VA-based analytical approach
was elaborated, which showed good agreement with numerical results. A
noteworthy feature of the SBB in the asymmetric model is a possibility of
hysteresis in a broad region, while in the symmetric system the hysteresis
occurs in the narrow bistability region between the two bifurcation points,
as seen in Fig. \ref{fig2}. A systematic analysis of the MI of CW states in
the model of asymmetric nonlinear couplers was reported in Ref. \cite{India2}.

\begin{figure}[tbp]
\begin{center}
\includegraphics[height=10cm]{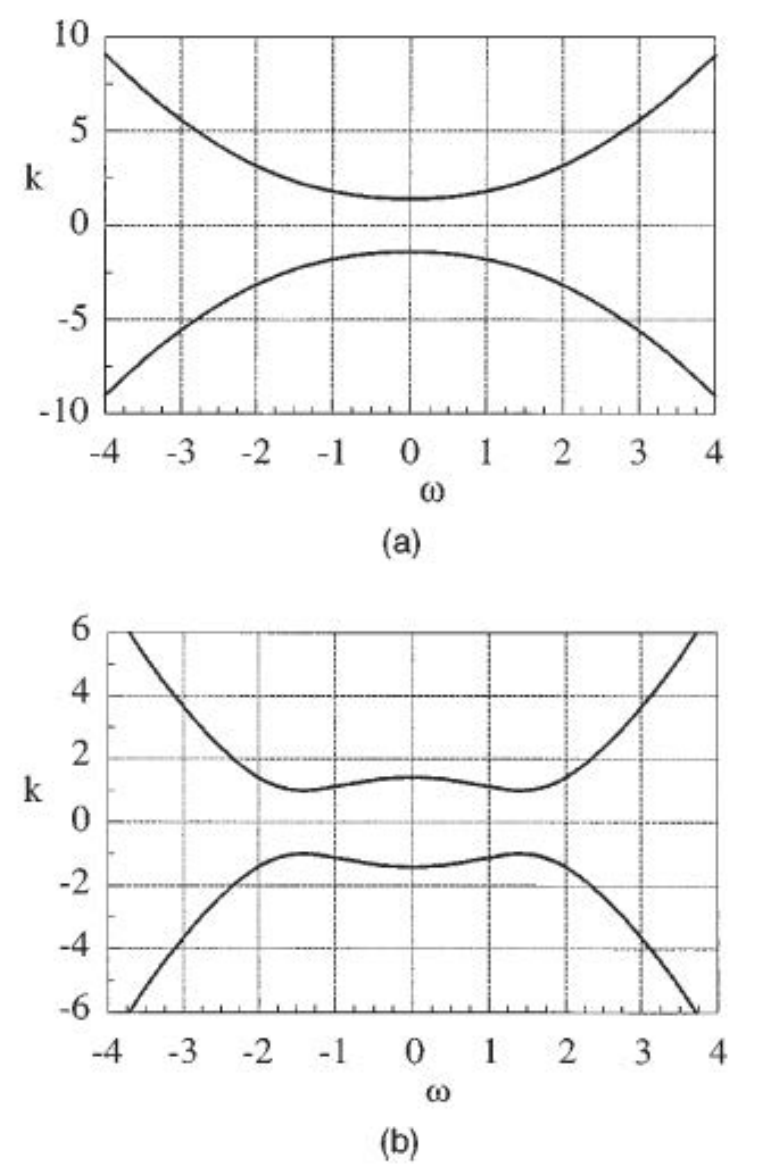}
\end{center}
\caption{Typical dispersion curves produced by Eq. (\protect\ref%
{asymmdispersion}) for the dual-core fiber with opposite signs of the GVD in
the cores, corresponding to Eqs. (\protect\ref{uasymm}) and (\protect\ref%
{vasymm}) with $\protect\delta =1$: (a) $q=-1$; (b) $q=+1$ (as per Ref.
\protect\cite{KM-1998}).}
\label{fig3}
\end{figure}

The most interesting version of the asymmetric model is one with $\delta >0$
in Eq. (\ref{vasymm}), i.e., with \emph{opposite} signs of the GVD \cite%
{KM-1998}. In this case, the substitution of $u,v\sim \exp \left(
ikz-i\omega \tau \right) $ in the linearized version of Eqs. (\ref{uasymm})
and (\ref{vasymm}) yields the respective dispersion relation,
\begin{equation}
k=\frac{1}{4}\left( \delta -1\right) \left( \omega ^{2}-2q\right) \pm \sqrt{%
\frac{1}{16}\left( \delta +1\right) ^{2}\left( \omega ^{2}-2q\right) ^{2}+1}.
\label{asymmdispersion}
\end{equation}%
Self-trapped states may exist, as \emph{gap solitons} (GSs), at values of
the propagation constant, $k$, that belong to the \emph{gap} in spectrum (%
\ref{asymmdispersion}), i.e., such that values of $\omega $ corresponding to
given $k$, as per Eq. (\ref{asymmdispersion}), are unphysical (imaginary of
complex). The gap always exists in the case of $\delta >0$, as seen from
typical examples of the spectra for negative and positive mismatch $q$,
which are displayed in Fig. \ref{fig3}. If the formal values of $\omega $ in
the gap are complex, GS's tails decay with oscillations, while for pure
imaginary $\omega $ they decay monotonously. In particular, it follows from
Eq. (\ref{asymmdispersion}) that, in subgap $0\leq k^{2}<4\delta /\left(
1+\delta \right) ^{2}$, the tails always decay with oscillations.

Solutions to Eqs. (\ref{uasymm}) and (\ref{vasymm}) for stationary GSs are
sought for as $u\left( z,\tau \right) =U(\tau )\exp \left( ikz\right) $, $%
v\left( z,\tau \right) =V(\tau )\exp \left( ikz\right) $, with real $U$ and $%
V$ determined by equations
\begin{gather}
(q-k)U+\frac{1}{2}\frac{d^{2}U}{d\tau ^{2}}+U^{3}+V=0,\;  \notag \\
-(\delta q+k)V-\frac{1}{2}\delta \frac{d^{2}V}{d\tau ^{2}}+V^{3}+U=0.
\label{asymmreal}
\end{gather}%
Approximate solutions to Eqs. (\ref{asymmreal}) can be constructed by means
of the VA, using the Gaussian ansatz

\begin{equation}
U=A\;\exp \left( -\tau ^{2}/2a^{2}\right) ,~V=B\;\exp \left( -\tau
^{2}/2b^{2}\right) .  \label{asymmansatz}
\end{equation}%
Energies of the two components of the soliton corresponding to this ansatz
are
\begin{equation}
E_{u}\equiv \int_{-\infty }^{+\infty }\left\vert U(\tau )\right\vert ^{2}dt=%
\sqrt{\pi }A^{2}a,~E_{v}\equiv \int_{-\infty }^{+\infty }\left\vert V(\tau
)\right\vert ^{2}dt=\sqrt{\pi }B^{2}b,  \label{asymmenergy}
\end{equation}%
with the net energy $E\equiv E_{u}+E_{v}$.

Elimination of amplitudes $A$ and $B$ from the resulting system of
variational equations leads to coupled equations for the widths $a$ and $b$,
\begin{eqnarray}
\left[ 3-4(k-q)a^{2}\right] \left[ 3\delta +4(k+\delta q)b^{2}\right]
&=&32(ab)^{3}\left( b^{2}-3a^{2}\right) \left( 3b^{2}-a^{2}\right) \left(
a^{2}+b^{2}\right) ^{-3},  \label{aasymm} \\
\frac{\left[ 3-4(k-q)a^{2}\right] \left( 3a^{2}-b^{2}\right) ^{2}}{\left[
3\delta +4(k+\delta q)b^{2}\right] \left( 3b^{2}-a^{2}\right) ^{2}} &=&\frac{%
a^{3}\left[ \delta \left( 3a^{2}+b^{2}\right) +4(k+\delta q)b^{2}\left(
b^{2}-a^{2}\right) \right] }{b^{3}\left[ \left( 3b^{2}+a^{2}\right)
+4(k-q)a^{2}\left( b^{2}-a^{2}\right) \right] }.  \label{basymm}
\end{eqnarray}%
These equations can be solved numerically, to find $a$ and $b$ as functions
of propagation constant $k$ and parameters $\delta $ and $q$.

\begin{figure}[tbp]
\begin{center}
\includegraphics[height=10cm]{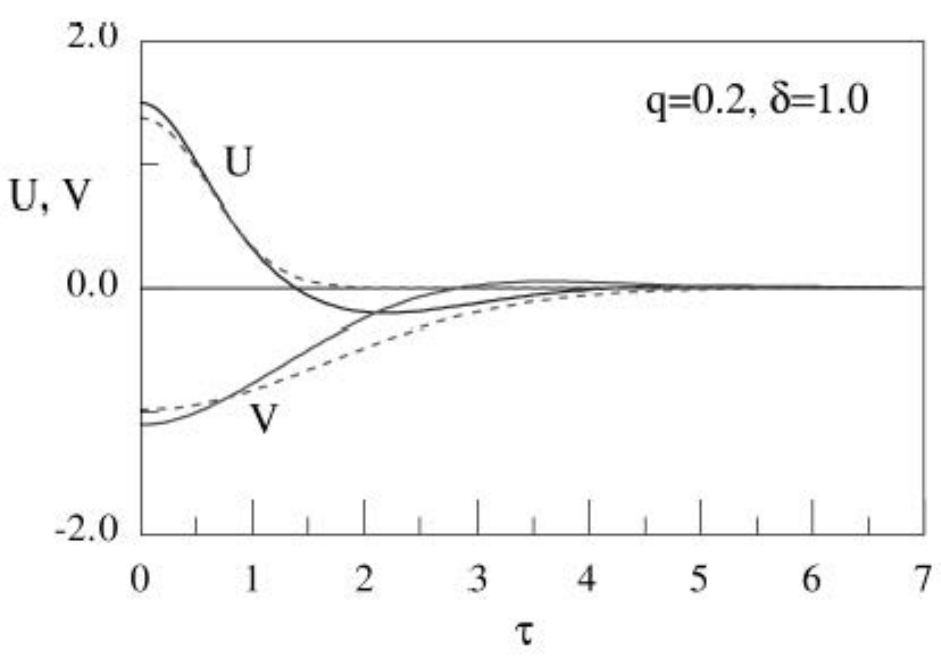}
\end{center}
\caption{A numerically found (solid lines) gap-soliton solution of Eqs. (%
\protect\ref{uasymm}) and (\protect\ref{vasymm}) with oscillating decaying
tails, and its VA-predicted counterpart (dashed lines), in the case of $%
\protect\delta =1$ (opposite GVD in the two cores) and $q=0.2$. The total
energy of the numerically found gap soliton is $E=2.734$ (as per Ref.
\protect\cite{KM-1998}).}
\label{fig4}
\end{figure}

The results reported in \cite{KM-1998} demonstrate that\ the GSs indeed
exist in a part of the available gap, and, in most cases, they are stable.
However, another part of the gap remains \emph{empty} (there are intervals
of $k$ in the gap where no soliton can be found). A noteworthy feature of
the GSs is that more than half of their net energy \emph{always} resides in
the normal-GVD component $v$, in spite of the obvious fact that the
normal-GVD core cannot, by itself, support any bright soliton. Further, a
typical GS predicted by the VA (see Fig. \ref{fig4}) has a narrower
component with a larger amplitude in the anomalous-GVD core, and a broader
component with a smaller amplitude in the normal-GVD one, see Fig. \ref{fig4}%
.

As is seen from Fig. \ref{fig4}, the VA generally correctly approximates the
soliton's core, but the simplest ansatz (\ref{asymmansatz}) does not take
into regard the fact that, as it mentioned above, the soliton's tails decay
with oscillations. The contribution of the tails also accounts for a
conspicuous difference of the energy share $E_{v}/E$ in the normal-GVD core
from the value predicted by the VA for the same net energy $E$: for example,
in the case shown in Fig. \ref{fig4}, the VA-predicted value is $%
E_{v}/E=0.585$, while its numerically found counterpart is $E_{v}/E=0.516$
(but it exceeds $1/2$, as stressed above).

\subsection{The coupler with separated nonlinearity and dispersion}

For better understanding of the light dynamics in strongly asymmetric
nonlinear couplers, it is relevant to consider the model of an extremely
asymmetric dual-core waveguide, in which the Kerr nonlinearity is carried by
one core, and the GVD is concentrated in the other \cite{Zafrany}. Such a
system, although looking \textquotedblleft exotic", can be created by means
of available technologies, by adjusting the zero-dispersion point of the
first core to the carrier wavelength of the optical signal, and using a
large effective cross-section area in the first core to suppress its
nonlinearity. The respective system of coupled equations is
\begin{gather}
iu_{z}+|u|^{2}u+v=0,  \label{Zu} \\
iv_{z}+qv+(D/2)v_{\tau \tau }+u=0,  \label{Zv}
\end{gather}%
cf. Eqs. (\ref{ucoupler}) and (%
\index{vcoupler}). Here the inter-core coupling coefficient is normalized to
be $1$, real parameter $q$, which may be positive or negative, is the
phase-velocity mismatch between the cores, and $D$ is the GVD coefficient,
that we may be scaled to be $+1$ or $-1$, which corresponds to the anomalous
or normal GVD, respectively. Group-velocity terms, such as $ic_{1}u_{\tau }$
in Eq. (\ref{Zu}) and $ic_{2}v_{\tau }$ in Eq. (\ref{Zv}), with some real
coefficients $c_{1}$ and $c_{2}$, can be removed: the former one by the
shift of the velocity of the references frame, $\tau \rightarrow \tau -c_{1}z
$, and the latter one by the phase transformation, $v\rightarrow v\exp
\left( ic_{2}\tau /D\right) $. Therefore, these terms are not included.

Looking for a solution to the linearized version of Eqs. (\ref{Zu}) and (\ref%
{Zv}) in the usual form, $\left\{ u,v\right\} \sim \exp \left( ikx-i\omega
\tau \right) $ with real $\omega $, one arrives at the dispersion relation,%
\begin{equation}
k=-%
\frac{1}{2}\left( \frac{1}{2}D\omega ^{2}-q\right) \pm \sqrt{\frac{1}{4}%
\left( \frac{1}{2}D\omega ^{2}-q\right) ^{2}+1}.  \label{Zk}
\end{equation}%
Straightforward consideration of the spectrum defined by this expression
demonstrates that, in the case of the anomalous GVD ($D=+1$), it gives rise
to \textit{finite} and \textit{semi-infinite gaps},%
\begin{equation}
-\frac{1}{2}\left( \sqrt{4+q^{2}}-q\right) <k<0;~\frac{1}{2}\left( \sqrt{%
4+q^{2}}+q\right) <k<\infty ,  \label{gapsD=+1}
\end{equation}%
and in the case of the normal GVD ($D=-1$), the \textit{semi-infinite} and
\textit{finite gaps} are%
\begin{equation}
-\infty <k<~\frac{1}{2}\left( \sqrt{4+q^{2}}-q\right) ;~0<k<\frac{1}{2}%
\left( \sqrt{4+q^{2}}+q\right) .  \label{gapsD=-1}
\end{equation}%
Note that both gaps (\ref{gapsD=+1}) are broader than their counterparts (%
\ref{gapsD=-1}) at $q>0$, and vice versa at $q<0$.

\begin{figure}[tbp]
\begin{center}
\includegraphics[height=12cm]{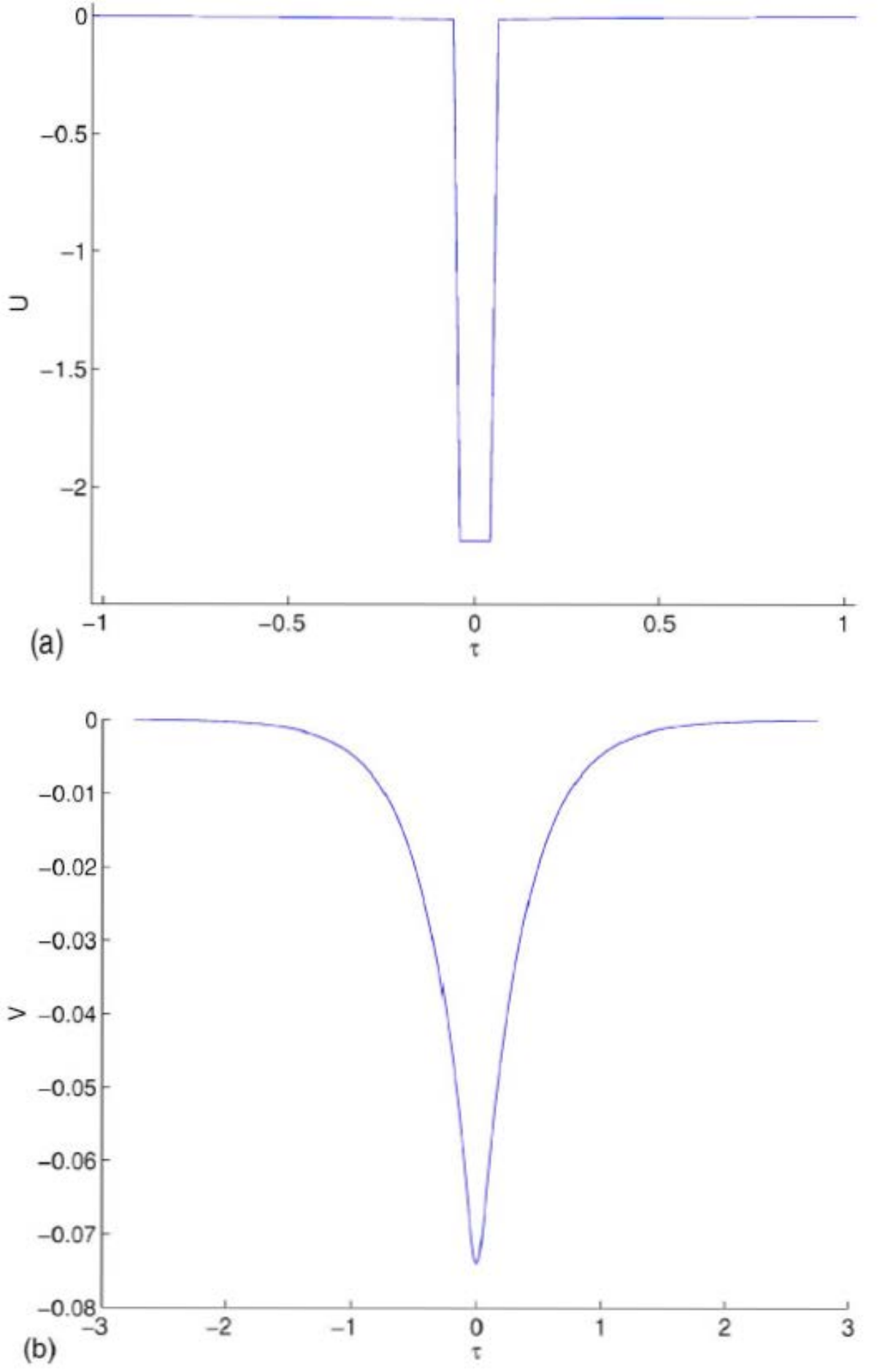}
\end{center}
\caption{An example of a stable soliton generated by the two-core system
with separated nonlinearity and dispersion, based on Eqs. (\protect\ref{Zu})
and (\protect\ref{Zv}), with parameters $D=1$ (the anomalous sign of the
GVD) and mismatch $q=0.8$. The propagation constant corresponding to this
soliton is $k=5$, which places it in the semi-infinite gap, see Eq. (\protect
\ref{gapsD=+1}). Panels (a) and (b) display, respectively, the nonlinear-
and dispersive-mode components of the soliton. }
\label{fig1add}
\end{figure}

Equation (\ref{Zk}) can be inverted, to yield $\omega ^{2}=2\left( Dk\right)
^{-1}\left( 1+qk-k^{2}\right) $. This relation implies that, inside both the
finite and semi-infinite gaps, $\omega ^{2}$ takes real negative values,
suggesting a possibility to find exponentially localized solitons in both
gaps. To realize this possibility, soliton solutions of Eqs. (\ref{Zu}) and (%
\ref{Zv}) were looked for as $\left\{ u,v\right\} =\exp \left( ikz\right)
\left\{ U(\tau ),V(\tau )\right\} $. In the case of anomalous GVD, $D=+1$,
it was thus found that the semi-infinite gap is \emph{completely filled} by
\emph{stable} solitons, while the finite bandgap remains completely empty.
This result does not depend on the magnitude and sign of the mismatch
parameter, $q$, in Eq. (\ref{Zv}). A typical example of the stable solitons
found in the semi-infinite gap is shown in Fig. \ref{fig1add}. Naturally,
the shape of the soliton in the dispersive mode ($V$) is much smoother than
in the nonlinear one ($U$). Nevertheless, the shapes of both components are
strictly smooth; in particular, there is no true cusp at the tip of the
dispersive one.

\begin{figure}[tbp]
\begin{center}
\includegraphics[height=12cm]{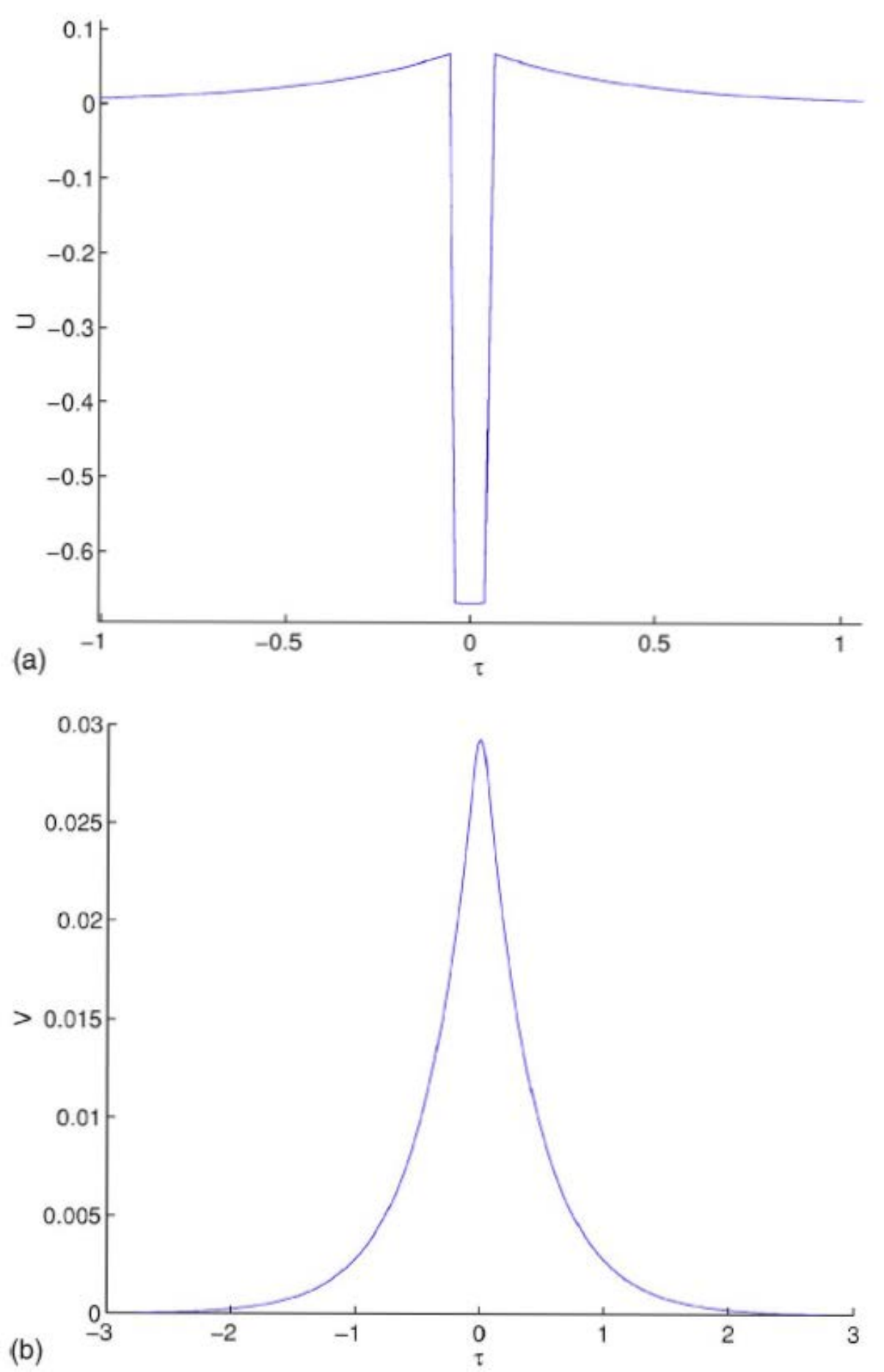}
\end{center}
\caption{The same as in Fig. \protect\ref{fig1add}, but for parameters $D=-1$
(the normal sign of the GVD) and mismatch $q=0.8$. The propagation constant
of this stable soliton is $k=0.4$, placing it in the finite bandgap, see Eq.
(\protect\ref{gapsD=-1}).}
\label{fig2add}
\end{figure}

The fact that the anomalous GVD supports stable solitons in the
semi-infinite gap of the present system is not surprising, as the situation
seems qualitatively similar to what is commonly known for the usual NLSE,
even if the shape of the solitons is very different from that in the NLSE,
see Fig. \ref{fig1add}. More unexpected is the situation in the case of the
normal GVD, $D=-1$ in Eq. (\ref{Zv}). As found in Ref. \cite{Zafrany}, in
this case the semi-infinite gap remains empty, but the \emph{finite} one is
\emph{completely filled} by \emph{stable GSs}, see a typical example in Fig. %
\ref{fig2add}.

It is relevant to mention that Eqs. (\ref{Zu}) and (\ref{Zv}) are not
Galilean invariant. In accordance with this, it was not possible to create
moving solitons in the framework of this system.

The analysis was also extended for the case when the nonlinear mode has weak
residual GVD, of either sign \cite{Zafrany}. Still earlier, a similar model
was considered in Ref. \cite{Javid-semilinear}, which introduced a two-core
system with the nonlinearity in one core, and a linear BG in the other. That
system creates a rather complex spectral structure, featuring three
bandgaps, and a complex family of soliton solutions, including the so-called
\textit{embedded} solitons, which, under special conditions, may exists in
(be \textit{embedded into})  spectral bands filled by linear waves, where,
generically, solitons cannot exist \cite{embedded}.

\subsection{Two polarizations of light in the dual-core fiber}

A relevant extension of the model of the nonlinear coupler takes into regard
two linear polarizations of light in each core. In this case, Eqs. (\ref%
{uasymm}) and (\ref{vasymm}) are replaced by a system of four equations,
\begin{equation}
\begin{array}{l}
i\left( u_{1}\right) _{z}+\frac{1}{2}\left( u_{1}\right) _{\tau \tau
}+\,(|u_{1}|^{2}+\frac{2}{3}|v_{1}|^{2})u_{1}+u_{2}=0,\vspace{0.2cm} \\
i\left( v_{1}\right) _{z}+\frac{1}{2}\left( v_{1}\right) _{\tau \tau
}+\,(|v_{1}|^{2}+\frac{2}{3}|u_{1}|^{2})v_{1}+v_{2}=0\vspace{0.2cm}, \\
i\left( u_{2}\right) _{z}+\frac{1}{2}\left( u_{2}\right) _{\tau \tau
}+\,(|u_{2}|^{2}+\frac{2}{3}|v_{2}|^{2})u_{2}+u_{1}=0,\vspace{0.2cm} \\
i\left( v_{2}\right) _{z}+\frac{1}{2}\left( v_{2}\right) _{\tau \tau
}+\,(|v_{2}|^{2}+\frac{2}{3}|u_{2}|^{2})v_{2}+v_{1}=0,%
\end{array}
\label{2x2}
\end{equation}%
where fields $u$ and $v$ represent the two linear polarizations, the
subscripts $1$ and $2$ label the cores, and the coupling coefficient is
scaled to be $K\equiv 1$. In the case of circular polarizations (rather than
linear ones), the XPM\ (cross-phase-modulation) coefficient $2/3$ in Eq. (%
\ref{2x2}) is replaced by $2$.

\begin{figure}[tbp]
\begin{center}
\includegraphics[height=12cm]{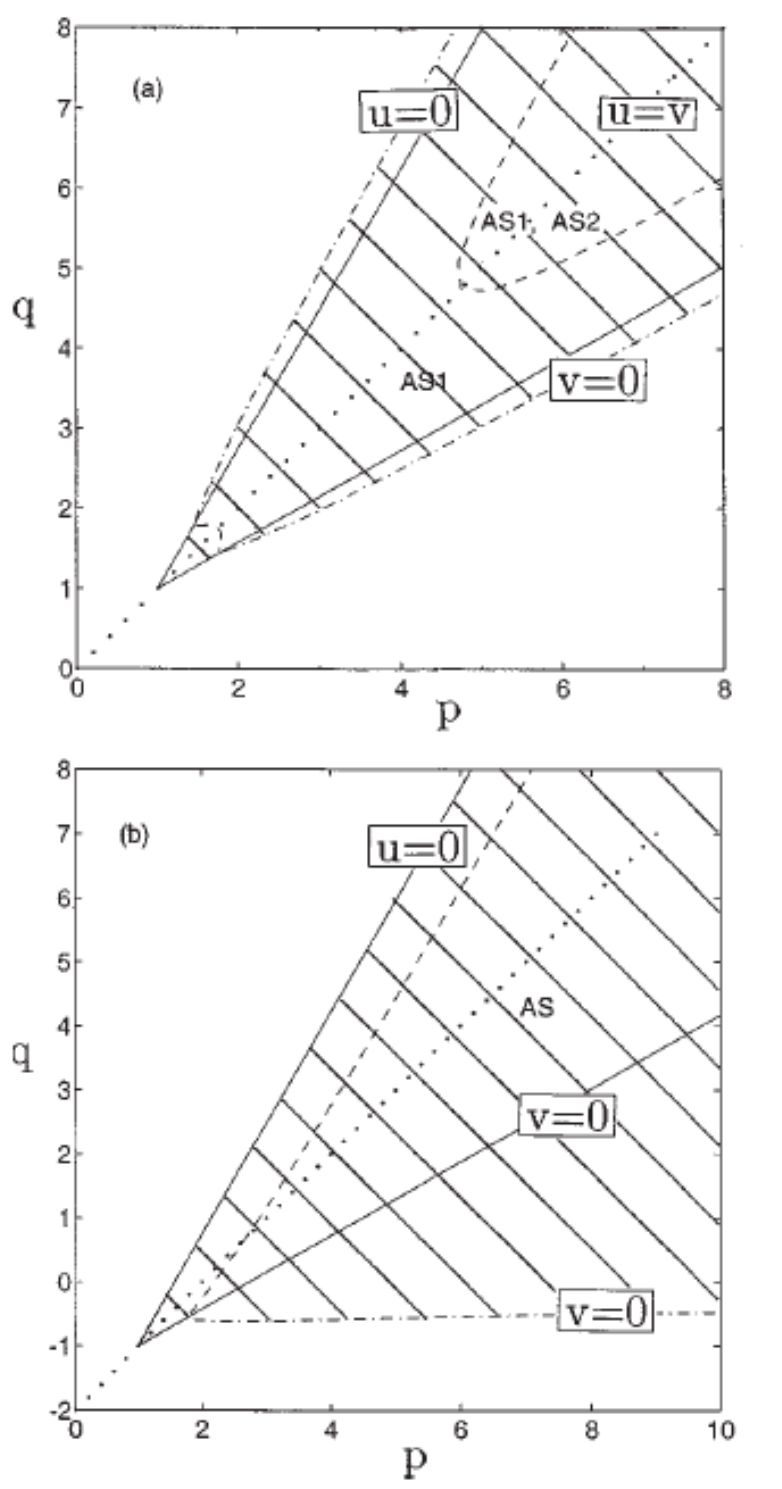}
\end{center}
\caption{Regions of existence of the symmetric and two types of asymmetric
(stable, AS1, and unstable, AS2) solitons in the plane ($p,q$) of two
propagation constants of four-component solitons (\protect\ref{4ansatz}), in
the model (\protect\ref{2x2}) of the dual-core-fiber system carrying two
linear polarizations of light. Symbols $u=0$, $v=0$, and $u=v$ refer to
particular solutions with a single polarization and equal amplitudes of the
two polarizations, respectively.}
\label{fig5}
\end{figure}

Four-component soliton solutions to Eqs. (\ref{2x2}) can be looked for by
means of the VA based on the Gaussian ansatz,
\begin{equation}
u_{1,2}(z,\tau )=A_{1,2}\exp \left( ipz-a^{2}\tau ^{2}/2\right)
,\,v_{1,2}(z,\tau )=B_{1,2}\exp \left( iqz-b^{2}\tau ^{2}/2\right) ,
\label{4ansatz}
\end{equation}%
with mutually independent real propagation constants $p$ and $q$. Existence
regions for all the solutions in the $\,(p,q)$ plane, produced by the VA for
symmetric and asymmetric solitons (the asymmetry is again realized with
respect to the two mutually symmetric cores) are displayed in Fig. \ref{fig5}%
, in the most essential case when the signs of amplitudes $A_{1,2}$ and $%
B_{1,2}$ in each polarization coincide (otherwise, all the solitons are
unstable). Outside the shaded area in Fig. \ref{fig5}, there exist only
solutions with a single polarization (i.e., with either $v_{1,2}=0$ or $%
u_{1,2}=0$), which were considered above. In particular, at the
dashed-dotted borders of the shaded area, asymmetric four-component solitons
(denoted by symbol AS1 in Fig. \ref{fig5}) carry over into the two-component
asymmetric solitons of the single-polarization system. The symmetric
solitons exist inside the sector bounded by straight continuous lines. The
SBB, which gives rise to the asymmetric solitons AS1 and destabilizes the
symmetric ones, takes place along the short-dashed curve in the left lower
corner of the shaded area.

There is an additional asymmetric soliton (AS2 in Fig. \ref{fig5}) in the
inner area confined by the dashed curve. Thus, the total number of soliton
solutions changes, as one crosses the bifurcation\ curves in Fig. \ref{fig5}
from left to right, from $1$ to $3$ to $5$. However, soliton AS2 is
generated from the symmetric one by an additional SBB, which takes place
\emph{after} the symmetric soliton has already been destabilized by the
bifurcation that gives rise to asymmetric soliton AS1. For this reason,
soliton AS2 is always unstable, while the primary asymmetric one AS1 is
stable. Further details concerning the stability of different solitons in
this model can be found in \cite{LK}.

\subsection{Solitons in linearly coupled fiber Bragg gratings (BGs)}

In the systems described by the single or coupled NLSEs, the
second-derivative terms account for the intrinsic GVD of the fiber or
waveguide. On the contrary to this, strong \emph{artificial dispersion} can
be induced by a BG, i.e., a permanent periodic modulation of the refractive
index written along the fiber (usually, the modulation is created in the
fiber's cladding), the modulation period being equal to half the wavelength
of the propagating light. The nonlinear optical fiber carrying the BG is
adequately described by the system of coupled-mode equations for amplitudes $%
u(x,t)$ and $v\left( x,t\right) $ of the right- and left-traveling waves
\cite{deS-S,Ace}:
\begin{eqnarray}
iu_{t}+iu_{x}+\left[ (1/2)|u|^{2}+|v|^{2}\right] u+v &=&0,  \label{uBragg} \\
iv_{t}-iv_{x}+\left[ |u|^{2}+(1/2)|v|^{2}\right] v+u &=&0.  \label{vBragg}
\end{eqnarray}%
Here, the speed of light in the fiber's material is scaled to be $1$, as
well as the linear-coupling constant, that accounts for mutual conversion of
the right- and left traveling waves due to the resonant reflection of light
on the BG. The ratio of the XPM and SPM (self-phase-modulation) coefficients
in Eqs. (\ref{uBragg}) and (\ref{vBragg}), $2:1$, is the usual feature of
the Kerr nonlinearity.

The dispersion relation of the linearized version of Eqs. (\ref{uBragg}) and
(\ref{vBragg}) is $\omega ^{2}=1+k^{2}$, hence the existence of GSs (alias
BG solitons) with frequencies belonging to the corresponding spectral
bandgap, $-1<\omega <+1$, may be expected. Indeed, although the system of
Eqs. (\ref{uBragg}) and (\ref{vBragg}) is not integrable, it has a family of
exact soliton solutions \cite{Russian,AW,CJ}, which contains two nontrivial
parameters, \textit{viz}., amplitude $Q,$ which takes values $0<Q<\pi $, and
velocity $c$, which belongs to interval $-1<c<+1$. In particular, the
solution for the quiescent solitons ($c=0$) is
\begin{equation}
\begin{array}{rl}
u= & \sqrt{2/3}\left( \sin Q\right) ~\mathrm{sech}\left( x\sin Q-\frac{1}{2}%
iQ\right) \cdot \exp \left( -it\cos Q\right) , \\
v= & -\sqrt{2/3}\left( \sin Q\right) \,\mathrm{sech}\left( x\sin Q+\frac{1}{2%
}iQ\right) \cdot \exp \left( -it\cos Q\right) ,%
\end{array}
\label{solBragg}
\end{equation}%
where frequencies $\omega _{\mathrm{sol}}\equiv \cos Q$ precisely fill the
entire gap, while $Q$ varies between $0$ and $\pi $. Stability of the BG
solitons was investigated too, the result being that they are stable,
roughly, in a half of the bandgap, namely, at $0<Q<Q_{\mathrm{cr}}\approx
1.01\cdot \left( \pi /2\right) $ \cite{Rich,Barash,Rome}.

A natural generalization of the fiber BG is a system of two parallel-coupled
cores with identical gratings written on both of them \cite{Bragg}. The
respective system of four coupled equations can be cast in the following
normalized form, cf. Eqs. (\ref{uBragg}), (\ref{vBragg}) for the single-core
BG fiber, and Eqs. (\ref{ucoupler}), (\ref{vcoupler}) for the dual-core
fiber without the BG:
\begin{eqnarray}
iu_{1t}+iu_{1x}+(\frac{1}{2}|u_{1}|^{2}+|v_{1}|^{2})u_{1}+v_{1}+\lambda
u_{2} &=&0,  \label{bu1} \\
iv_{1t}-iv_{1x}+(\frac{1}{2}|v_{1}|^{2}+|u_{1}|^{2})v_{1}+u_{1}+\lambda
v_{2} &=&0\,,  \label{bv1} \\
iu_{2t}+iu_{2x}+(\frac{1}{2}|u_{2}|^{2}+|v_{2}|^{2})u_{2}+v_{2}+\lambda
u_{1} &=&0\,,  \label{bu2} \\
iv_{2t}-iv_{2x}+(\frac{1}{2}|v_{2}|^{2}+|u_{2}|^{2})v_{2}+u_{2}+\lambda
v_{1} &=&0\,,  \label{bv2}
\end{eqnarray}%
where $\lambda $ is the coefficient of the linear coupling between the two
cores, which may be defined to be positive (unlike the models considered
above, it is not possible to fix $\lambda =1$ by means of rescaling, because
the scaling freedom has been already used to fix the Bragg-reflection
coefficient equal to $1$). The same model applies to the spatial-domain
propagation in two parallel-coupled planar waveguides which carry BGs in the
form of a system of parallel cores, in which case $t$ and $x$ play the roles
of the propagation distance and transverse coordinate, respectively, while
the paraxial diffraction in the waveguides is neglected.

The dispersion relation for system (\ref{bu1})-(\ref{bv2}) contains four
branches (taking into regard that $\omega $ may have two opposite signs):
\begin{equation}
\omega ^{2}=\lambda ^{2}+1+k^{2}\pm 2\lambda \sqrt{1+k^{2}}.  \label{MakDisp}
\end{equation}%
This spectrum has no gap in the case of strong inter-core coupling, $\lambda
>1$. For the weaker coupling, with $\lambda <1$, the bandgap exists:%
\begin{equation}
-\left( 1-\lambda \right) <\omega <+\left( 1-\lambda \right) .
\label{BG-gap}
\end{equation}

To populate the bandgap, solutions for zero-velocity GSs are looked for as
\begin{equation}
u_{1,2}=\exp \left( -i\omega t\right) \,U_{1,2}(x)\,,\;v_{1,2}=\exp \left(
-i\omega t\right) \,V_{1,2}(x)\,,  \label{statu}
\end{equation}%
where relation $V_{1,2}=-U_{1,2}^{\ast }$ may be imposed\ (in fact, the
exact GS solutions (\ref{solBragg}) in the single-core BG are subject to the
same constraint) . Substituting this in Eqs.~(\ref{bu1})-(\ref{bv2}) leads
to two coupled equations (instead of four):
\begin{eqnarray}
\omega U_{1}+i\frac{dU_{1}}{dx}+\frac{3}{2}|uU_{1}|^{2}U_{1}-U_{1}^{\ast
}+\lambda U_{2} &=&0\,,  \label{rdstu1} \\
\omega U_{2}+i\frac{dU_{2}}{dx}+\frac{3}{2}|U_{2}|^{2}U_{2}-U_{2}^{\ast
}+\lambda U_{1} &=&0\,.  \label{rdstu2}
\end{eqnarray}%
Stationary equations~(\ref{rdstu1}) and (\ref{rdstu2}) can be derived from
their own Lagrangian, with density
\begin{gather}
\mathcal{L}=\omega (U_{1}U_{1}^{\ast }+U_{2}U_{2}^{\ast })+\frac{i}{2}\left(
\frac{dU_{1}}{dx}U_{1}^{\ast }-\frac{dU_{1}^{\ast }}{dx}U_{1}+\frac{dU_{2}}{%
dx}U_{2}^{\ast }-\frac{dU_{2}^{\ast }}{dx}U_{2}\right)   \notag \\
+\frac{3}{4}(|U_{1}|^{4}+|U_{2}|^{4})-\frac{1}{2}(U_{1}^{2}+U_{1}^{\ast
2}+U_{2}^{2}+U_{2}^{\ast 2})+\lambda (U_{1}U_{2}^{\ast }+U_{1}^{\ast
}U_{2})\,.  \label{Ldensity-BG}
\end{gather}%
Then, the following ansatz may be adopted for the \emph{complex} soliton\
solution sought for:
\begin{equation}
U_{1,2}=A_{1,2}\,\mathrm{sech}\,(\mu x)\,+iB_{1,2}\,\mathrm{sinh}(\,\mu x)\,%
\mathrm{sech}\,^{2}(\mu x)\,,  \label{nu2r}
\end{equation}%
with real $A_{1,2}$, $B_{1,2}$, and $\mu $. The integration of Lagrangian
density (\ref{Ldensity-BG}) with this ansatz and subsequent application of
the variational procedure gives rise to the following system of equations:
\begin{eqnarray}
3\lambda A_{2,1}-3(1-\omega )A_{1,2}+3A_{1,2}^{3}+\frac{3}{5}%
A_{1,2}B_{1,2}^{2}-\mu B_{1,2} &=&0\,,  \label{MakA} \\
\lambda B_{2,1}+\frac{3}{2}B_{1,2}-3.\,\allowbreak 857B_{1,2}^{3}+\frac{3}{5}%
A_{1,2}^{2}B_{1,2}-\mu A_{1,2} &=&0\,,  \label{MakB}
\end{eqnarray}%
\begin{gather}
2\omega (A_{1}^{2}+A_{2}^{2})+\frac{2\omega }{3}%
(B_{1}^{2}+B_{2}^{2})+(A_{1}^{4}+A_{2}^{4})-1.\,\allowbreak
285\,7(B_{1}^{4}+B_{2}^{4})  \notag \\
+\frac{2}{5}(A_{1}^{2}B_{1}^{2}+A_{2}^{2}B_{2}^{2})-2(A_{1}^{2}+A_{2}^{2})+%
\frac{2}{3}(B_{1}^{2}+B_{2}^{2})+4\lambda A_{1}A_{2}+\frac{4\lambda }{3}%
B_{1}B_{2}=0\,,  \label{varmu}
\end{gather}%
where numerical coefficients $3.\,\allowbreak 857$ and $1.285\,7$ are
defined by some integrals.

A general result, following\ from both a numerical solution of variational
equations (\ref{MakA}) - (\ref{varmu}) and direct numerical solution of Eqs.
(\ref{rdstu1}) and (\ref{rdstu2}), is that a symmetric mode, with $%
A_{1}^{2}=A_{2}^{2}$ and $B_{1}^{2}=B_{2}^{2}$, exists at all values of $%
\omega $ in the bandgap (\ref{BG-gap}), and it is the single soliton
solution if the coupling constant $\lambda $ is close enough to $1$, i.e.,
the bandgap (\ref{BG-gap}) is narrow. However, below a critical value of $%
\lambda $ (which depends on given $\omega $), the symmetric solution
undergoes a bifurcation, giving rise to three branches, one remaining
symmetric, while two new ones, which are mirror images to each other,
represent nontrivial \emph{asymmetric} solutions.

\begin{figure}[tbp]
\begin{center}
\includegraphics[height=10cm]{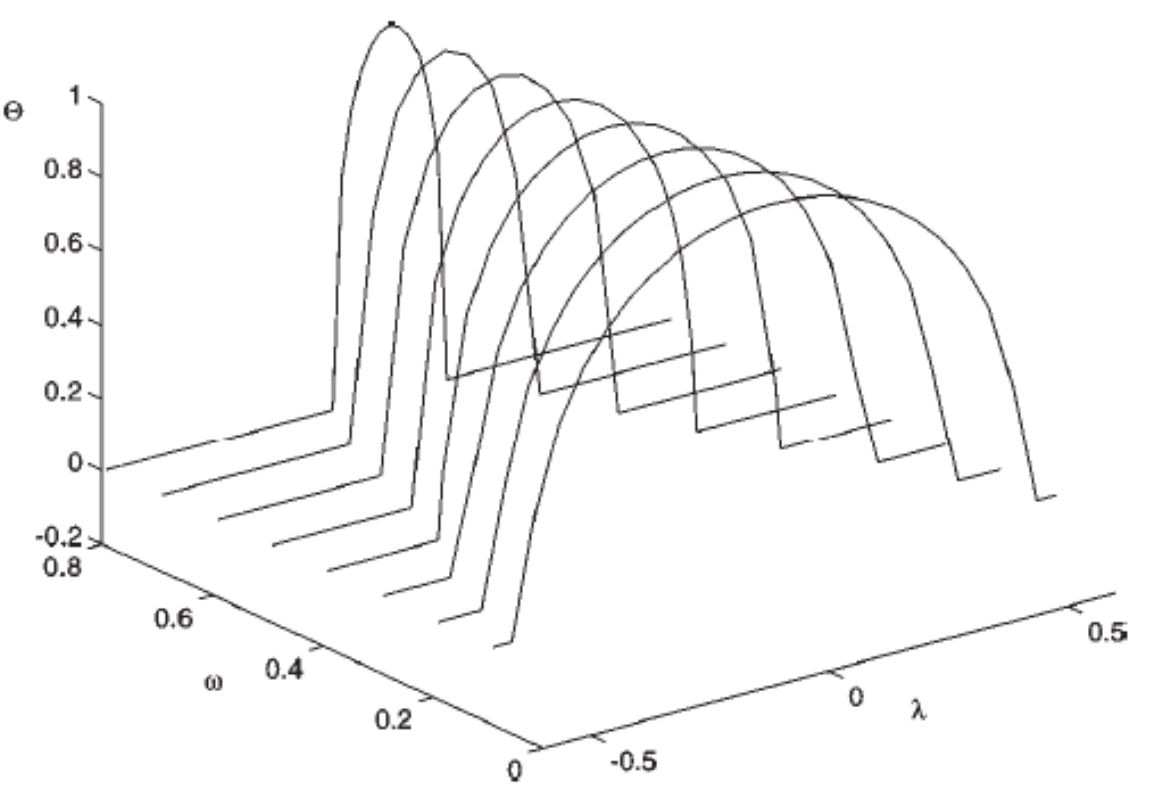}
\end{center}
\caption{The symmetry-breaking bifurcation diagram for zero-velocity gap
solitons in the model of the dual-core nonlinear optical fiber with
identical Bragg gratings written on both cores (as per Ref. \protect\cite%
{Bragg}).}
\label{fig6}
\end{figure}

The bifurcation can be conveniently displayed in terms of an effective
asymmetry parameter,
\begin{equation}
\Theta \equiv \left( U_{1m}^{2}-U_{2m}^{2}\right) /\left(
U_{1m}^{2}+U_{2m}^{2}\right) \,,  \label{Theta}
\end{equation}%
where $U_{1m}^{2}$ and $U_{2m}^{2}$ are peak powers (maxima of the squared
absolute values) of complex fields $U_{1,2}$ in the two cores (note its
difference from the asymmetry parameter (\ref{couplerasymmetry}), which was
defined in terms of integral energies, rather than peak powers). A complete
plot of the SBB for the GSs in the present system, i.e., $\Theta $ vs. $%
\omega $ and $\lambda $, is displayed in Fig.~\ref{fig6}. At $\lambda =0$,
when Eqs.~(\ref{rdstu1}) and (\ref{rdstu2}) decouple, the numerical solution
matches the exact solution (\ref{solBragg}) in one core, while the other
core is empty. Note the difference of this \emph{supercritical} (alias \emph{%
forward}) SBB from its weakly \emph{subcritical} (\emph{backward})
counterpart for the solitons in the nonlinear coupler without the BG, which
is shown above in Fig. \ref{fig2}.

\begin{figure}[tbp]
\begin{center}
\includegraphics[height=10cm]{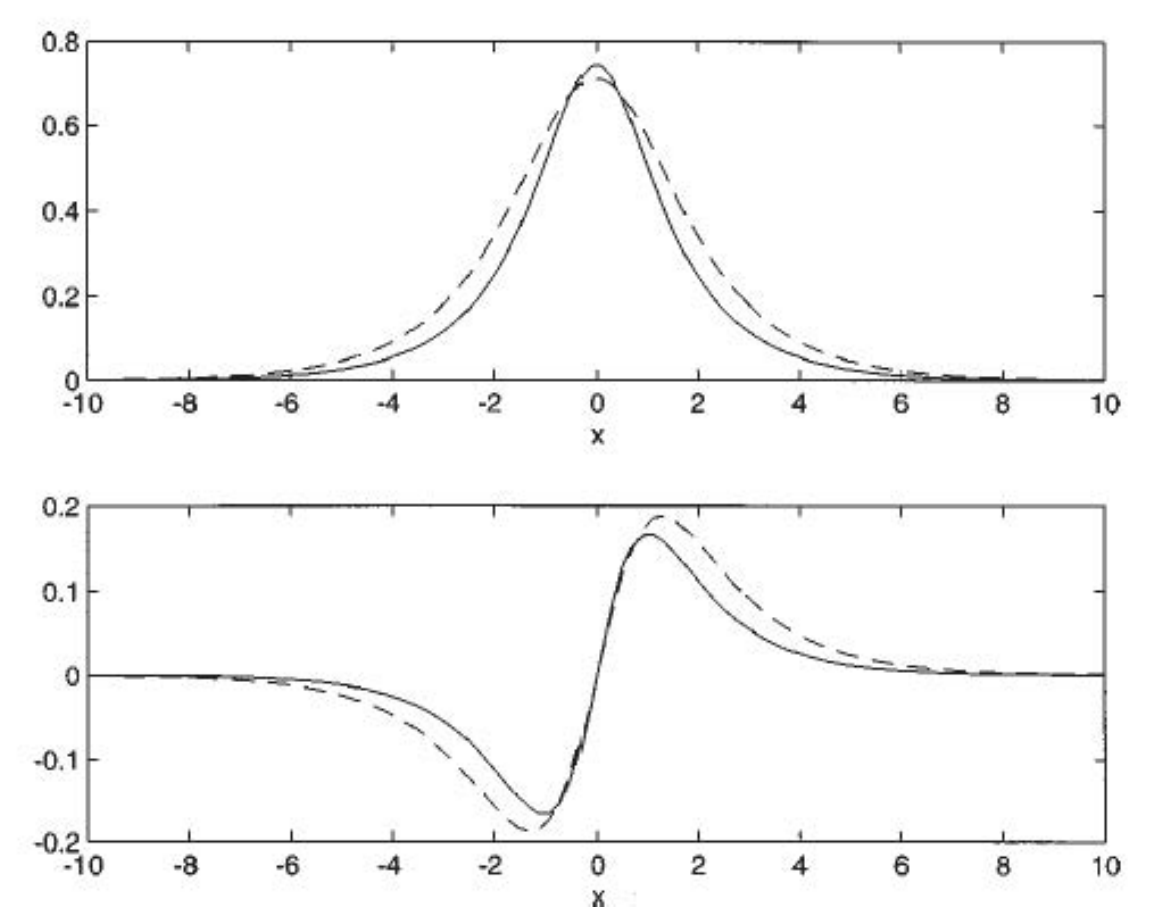}
\end{center}
\caption{Shapes of the larger component of the quiescent soliton, $U_{1}$,
in the dual-core Bragg grating (as per Ref. \protect\cite{Bragg}). The upper
and lower plots show the real and imaginary parts of $U_{1}$. Here, $\protect%
\omega =0.5$ and $\protect\lambda =0.2$.}
\label{fig7}
\end{figure}

The bifurcation diagram in Fig. \ref{fig6} was drawn using numerical results
obtained from the solution of Eqs. (\ref{rdstu1}) and (\ref{rdstu2}), but
its variational counterpart is very close to it, a relative discrepancy
between the VA-predicted and numerically exact values of $\lambda $, at
which the SBB takes place for fixed $\omega $, being $\,\lesssim 5\%$. To
directly illustrate the accuracy of the VA in the present case, comparison
between typical shapes of a stable asymmetric soliton, as obtained from the
full numerical solution, and as predicted by the VA, is presented in Fig. %
\ref{fig7}.

Direct numerical test of the stability of the symmetric and asymmetric
solitons in the present model has yielded results exactly corroborating what
may be expected: all the asymmetric solitons are stable whenever they exist,
while all the symmetric solitons, whenever they coexist with the asymmetric
ones, are unstable. However, all the symmetric solitons are\emph{\ }stable
prior to the bifurcation, where their asymmetric counterparts do not exist.

Lastly, it is relevant to mention that influence of a possible phase shift
between the BGs, written in the parallel-coupled cores, on four-component
GSs in this system was studied too \cite{Tsofe,Sukhorukov}. In that case,
the spontaneous (intrinsic) symmetry breaking is combined with the external
symmetry breaking imposed by the mismatch between the BGs.

\subsection{\emph{Bifurcation loops} for solitons in couplers with the
cubic-quintic (CQ)\ nonlinearity}

To conclude this section, it is relevant to briefly consider results
obtained for the coupler with the CQ nonlinearity, i.e., a combination of
competing self-focusing cubic and defocusing quintic terms in the respective
system of coupled NLSEs:
\begin{gather}
iu_{z}+u_{\tau \tau }+2|u|^{2}u-|u|^{4}u+\lambda v=0,  \label{uCQ} \\
iv_{z}+v_{\tau \tau }+2|v|^{2}v-|v|^{4}v+\lambda u=0,  \label{vCQ}
\end{gather}%
cf. the usual system of Eqs. (\ref{ucoupler}) and (\ref{vcoupler}) with the
cubic self-focusing (in both systems, the anomalous sign of the GVD is
assumed). By means of straightforward rescaling, the coefficients in front
of the nonlinear and dispersive terms may be fixed, without the loss of
generality, as written in Eqs. (\ref{uCQ}) and (\ref{vCQ}), while
coefficient $\lambda >0$ of the linear inter-core coupling remains a free
irreducible parameter. Note that the $\mathcal{PT}$-symmetric version of the
coupler with the CQ nonlinearity and solitons in it were considered too \cite%
{PT-CQ}.

The CQ combination of the competing nonlinearities, which is assumed in the
present system, occurs in various optical media. The realization which is
directly relevant to the fabrication of dual-core fibers is provided by
chalcogenide glasses \cite{CQglass1,CQglass2}.

The starting point of the analysis is a well-known exact soliton solution of
the single CQ NLS equation \cite{Pushkarov1,Pushkarov2}, to which Eqs. (\ref%
{u}) and (\ref{v}) reduce in the symmetric case:
\begin{eqnarray}
u &=&v=e^{ikz}U_{\mathrm{symm}}(\tau ),  \notag \\
U_{\mathrm{symm}}(\tau ) &=&\sqrt{\frac{2\left( k-\lambda \right) }{1+\sqrt{%
1-4\left( k-\lambda \right) /3}\cosh \left( 2\sqrt{k-\lambda }\tau \right) }}%
,  \label{Pushk}
\end{eqnarray}%
where the propagation constant $k$ takes values in the interval of $\lambda
<k<\frac{3}{4}+\lambda $. In the limit cases of $k=\lambda $ and
\begin{equation}
k=\frac{3}{4}+\lambda   \label{3/4}
\end{equation}%
this solution goes over, respectively, into the trivial zero solution, and
into the delocalized (continuous-wave, CW) state with a constant amplitude, $%
u=v=\sqrt{\frac{3}{2}}\exp \left( i\left( \frac{3}{4}+\lambda \right)
z\right) $. The energy of soliton (\ref{Pushk}), which is defined by the
same expression (\ref{E}) as above, is%
\begin{equation}
E_{\mathrm{symm}}=\frac{\sqrt{3}}{2}\ln \left( \frac{\sqrt{3}+2\sqrt{%
k-\lambda }}{\sqrt{3}-2\sqrt{k-\lambda }}\right) .  \label{Nsymm}
\end{equation}%
Naturally, it diverges in the limit corresponding to Eq. (\ref{3/4}).

Following the pattern of the above analysis, asymmetric stationary solitons
solutions to Eqs. (\ref{uCQ}) and (\ref{vCQ}) are looked for as
\begin{equation}
\left\{ u(z,\tau ),v(z,\tau )\right\} =e^{ikz}\left\{ U(\tau ),V(\tau
)\right\} .  \label{uv}
\end{equation}%
It can be proved \cite{Albuch} that only solutions with real functions $%
U(\tau )$ and $V(\tau )$ can be generated by the SBB from the symmetric
soliton (\ref{Pushk}), hence the substitution of expressions (\ref{uv}) with
real $U$ and $V$ in Eqs. (\ref{u}) and (\ref{v}) leads to a system
\begin{eqnarray}
\frac{d^{2}U}{d\tau ^{2}}-kU+\lambda V+2U^{3}-U^{5} &=&0,  \label{UCQ} \\
\frac{d^{2}V}{d\tau ^{2}}-kV+\lambda U+2V^{3}-V^{5} &=&0,  \label{VCQ}
\end{eqnarray}%
which was solved numerically, and the stability of the so found solitons was
then identified by dint of direct simulations \cite{Albuch}.

The numerical solution of Eqs. (\ref{UCQ}) and (\ref{VCQ}) produces a
sequence of bifurcation diagrams displayed in Figs. \ref{bifurcation1} and %
\ref{bifurcation2}. In these diagrams, the soliton's asymmetry parameter,
\begin{equation}
\epsilon \equiv \frac{U_{\max }^{2}-V_{\max }^{2}}{U_{\max }^{2}+V_{\max
}^{2}},  \label{epsilon}
\end{equation}%
where $U_{\max }^{2}$ and $V_{\max }^{2}$ are the peak powers of the two
components of the soliton, is shown versus its total energy. Note the
similarity of this definition of the asymmetry to that adopted above in the
form of Eq. (\ref{Theta}) for solitons in the dual-core BG.
\begin{figure}[tbp]
\includegraphics[width=3.3in]{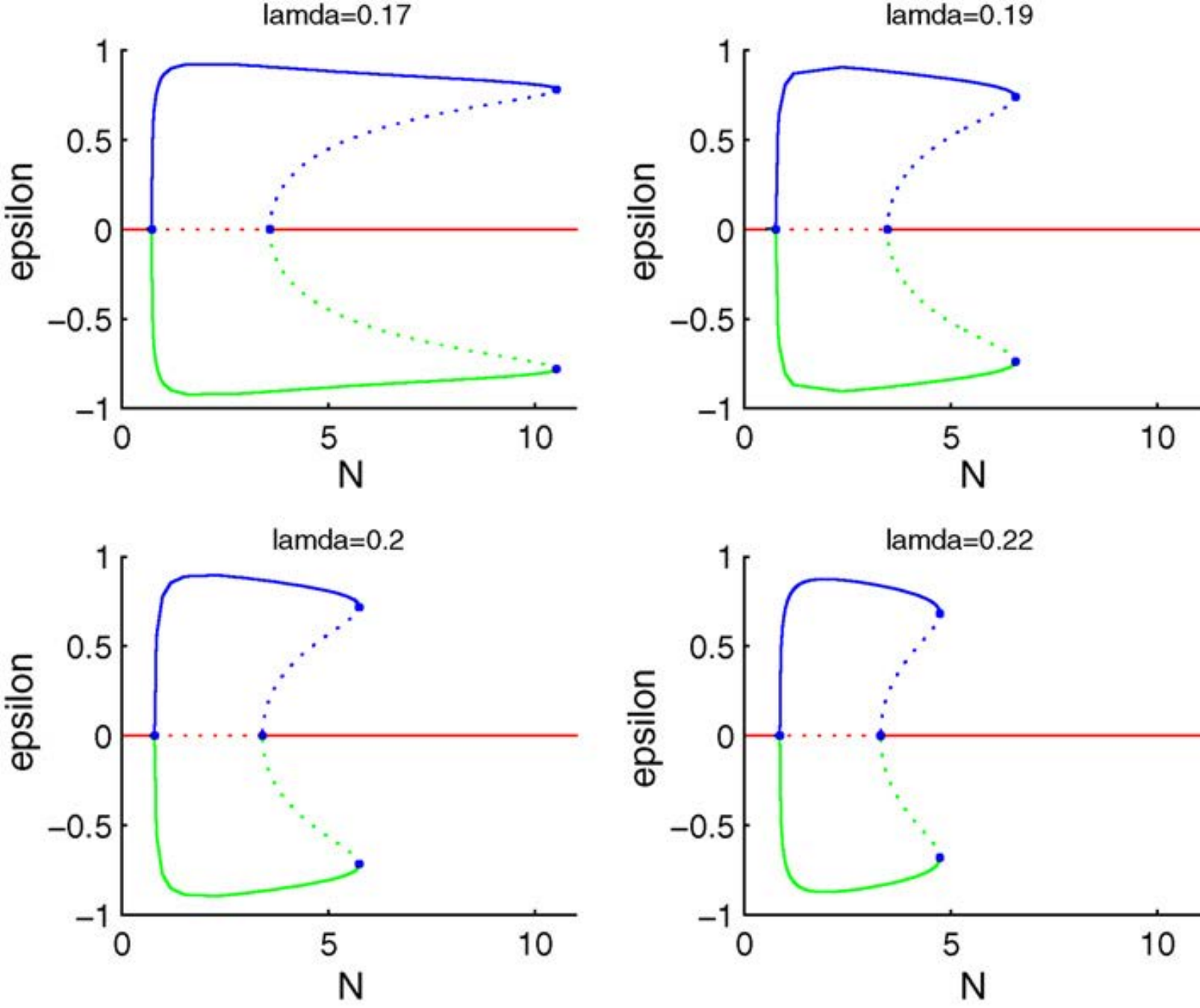}
\caption{A set of bifurcation diagrams for symmetric and asymmetric solitons
in the plane of the total energy, defined as in Eq. (\protect\ref{E}), but
denoted $N$ here (instead of $E$), and the asymmetry parameter (\protect\ref%
{epsilon}). The diagrams are produced by numerical solution of Eqs. (\protect
\ref{U}) and (\protect\ref{V}) with the cubic-quintic nonlinearity, at
different values of the linear-coupling constant, $\protect\lambda $. Stable
and unstable branches of the solutions are shown by solid and dashed curves,
respectively, and bold dots indicate bifurcation points (as per Ref.
\protect\cite{Albuch}).}
\label{bifurcation1}
\end{figure}
\begin{figure}[tbp]
\includegraphics[width=3.3in]{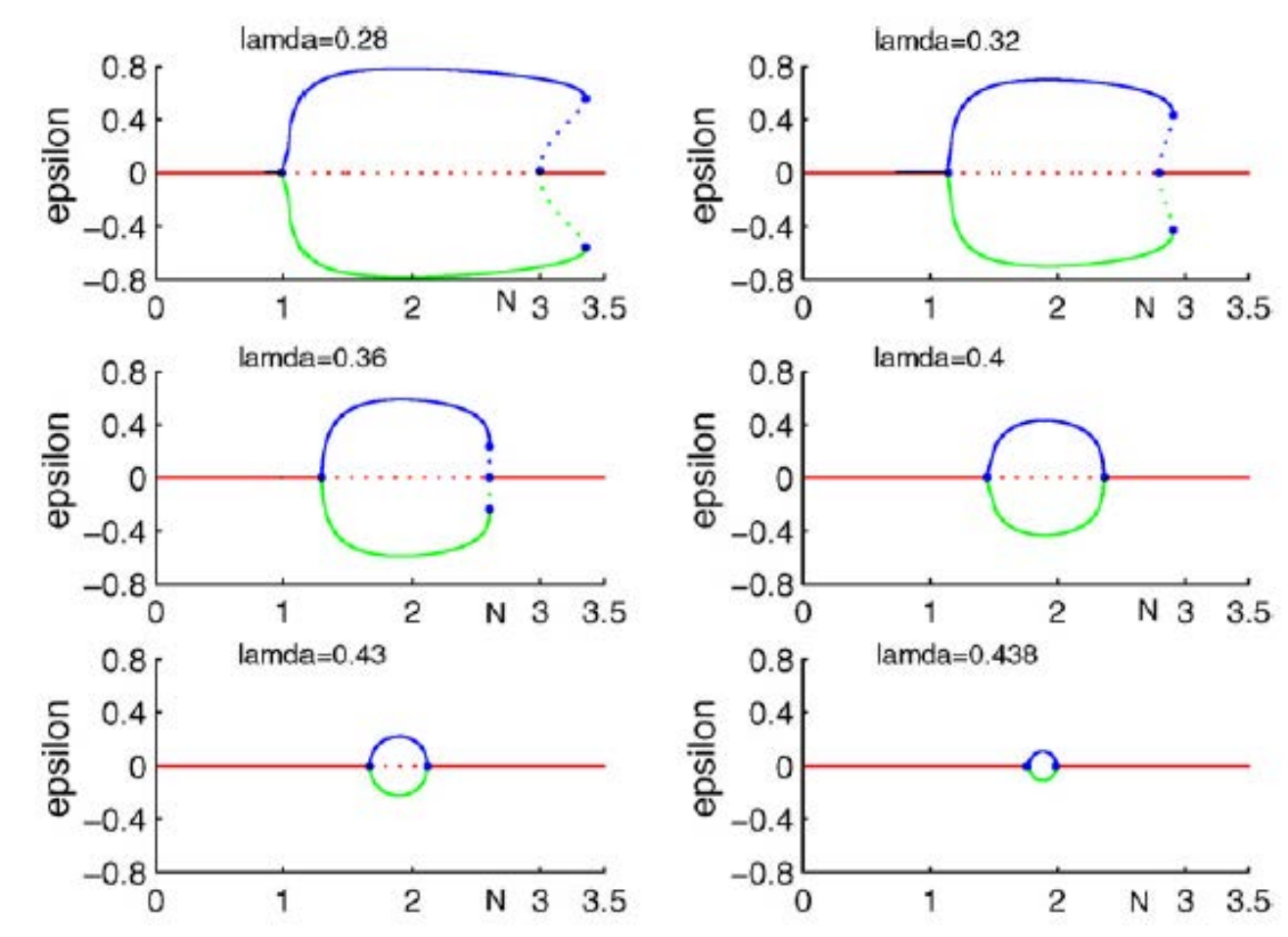}
\caption{Continuation of Fig. \protect\ref{bifurcation1} to larger values of
the coupling constant, $\protect\lambda $.}
\label{bifurcation2}
\end{figure}

\begin{figure}[tbp]
\includegraphics[width=3.3in]{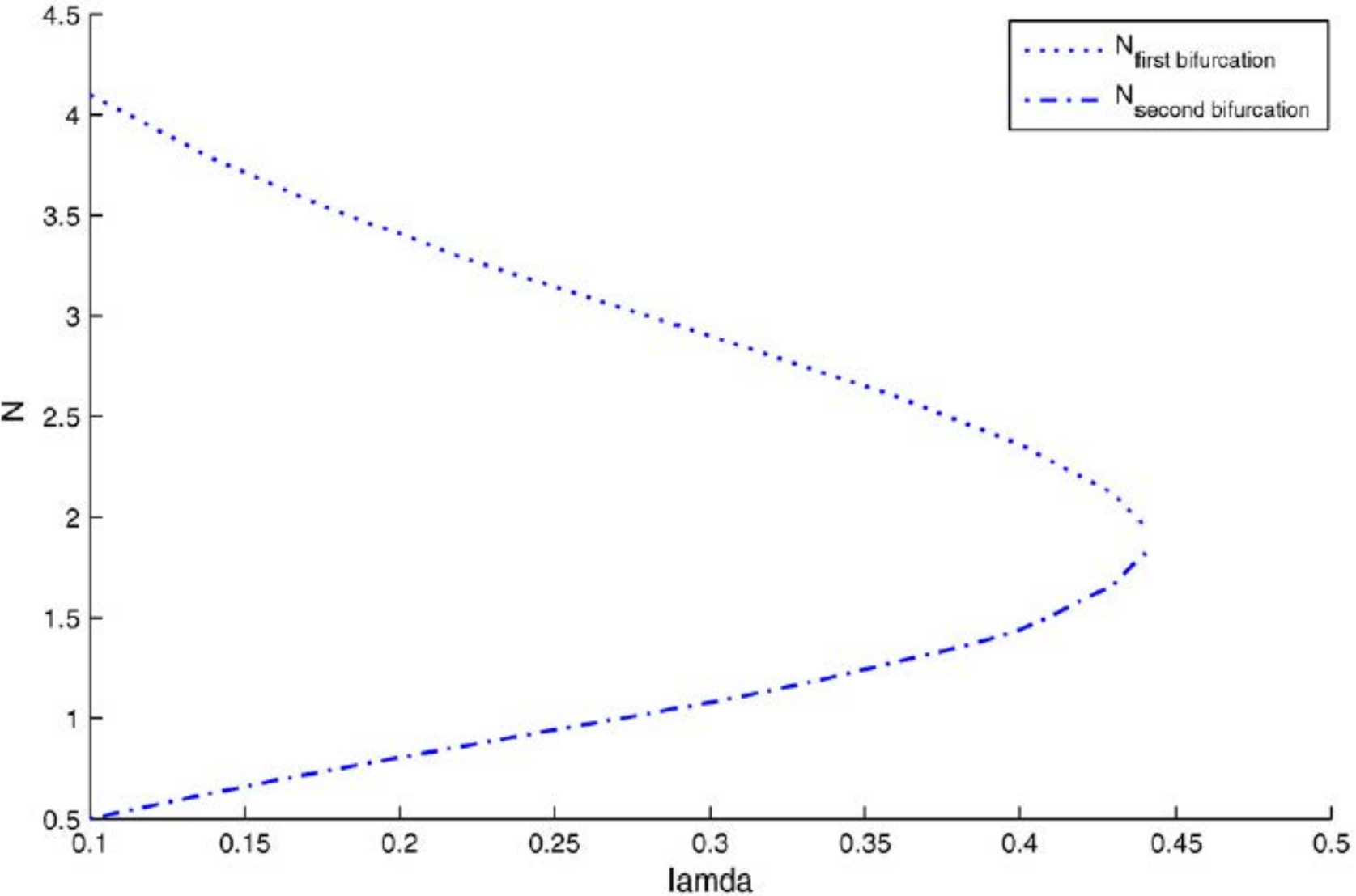}
\caption{Values of the energy of the symmetric soliton at which the direct
and reverse bifurcations occur in Figs. \protect\ref{bifurcation1} and
\protect\ref{bifurcation2}. The two curves merge and terminate at $\protect%
\lambda =\protect\lambda _{\max }\approx 0.44$.}
\label{bifurcation3}
\end{figure}

A remarkable peculiarity of the present system is the existence of the \emph{%
bifurcation loop}: as Figs. \ref{bifurcation1} and \ref{bifurcation2}
demonstrate, the \emph{direct} SBB, which occurs with the increase of the
energy, being driven, as above, by the cubic self-focusing, is followed, at
larger energies, by the \emph{reverse bifurcation}, which takes place when
the dominant nonlinearity becomes self-defocusing, represented by the
quintic terms in (\ref{UCQ}) and (\ref{VCQ}). The loop exists at $0<\lambda
\leq \lambda _{\max }\approx 0.44$. The direct bifurcation is seen to be
always supercritical, while the reverse one, which closes the loop, is
subcritical (giving rise to the bistability and concave shape of the loop,
on its right-hand side) up to $\lambda \approx 0.40$. In the interval of $%
0.40<\lambda <0.44$, the reverse bifurcation is supercritical, and the
(small) loop has a convex form. The picture of the bifurcations is
additionally illustrated by Fig. \ref{bifurcation3}, which displays the
energy of the symmetric soliton at points of the direct and reverse
bifurcations.

Stability and instability of different branches of the soliton solutions can
be anticipated on the basis of general principles of the bifurcation theory
\cite{bifurcations}: the symmetric solution becomes unstable after the
direct supercritical bifurcation, and asymmetric solutions emerge as stable
ones at this point; eventually, the reverse bifurcation restores the
stability of the symmetric solution. In the case when the reverse
bifurcation is subcritical and, accordingly, the bifurcation loop is concave
on its right side, two branches of asymmetric solutions meet at the turning
points, the branches which originate from the reverse-bifurcation point
being unstable. These expectations are fully borne out by direct numerical
simulations \cite{Albuch}. In particular, in the case when the bifurcation
loop has the concave shape, an unstable asymmetric soliton has a choice to
evolve into either a still more asymmetric one, or the symmetric soliton
(also stable). Numerical results clearly demonstrate that unstable
asymmetric solitons choose the former option, evolving into the \emph{more
asymmetric} counterparts.

\section{Dissipative solitons in dual-core fiber lasers}

\subsection{Introduction}

Experimental and theoretical studies of fiber lasers are, arguably, the
fastest developing area of the modern laser science \cite{doping}. A
commonly adopted model for the evolution of optical pulses in fiber lasers
is based on complex Ginzburg-Landau equations (CGLEs), which readily predict
formation of dissipative solitons, alias solitary pulses (SPs), in the
lasers \cite{Grelu}, due to the stable self-sustained balance of loss and
gain, the latter provided by the lasing mechanism (typically, stimulated
emission of photons by externally pumped ions of rear-earth metals which are
embedded as dopants into the fiber's silica \cite{doping}). Important
applications of the CGLEs are known in many other fields, including
hydrodynamics, plasmas, reaction-diffusion systems, etc., as well as other
areas of nonlinear optics \cite{Kramer,Encyclopedia}.

The CGLE of the simplest type is one with the linear dispersive gain and
cubic loss (which represents two-photon absorption), combined with the GVD
and Kerr nonlinearity. This equation readily produces an exact analytical
solution for SPs, in the form of the chirped hyperbolic secant \cite%
{Hocking,Lennart}, but they are unstable, for an obvious reason: the linear
gain destabilizes the zero background around the SP. The most
straightforward modification which makes the existence of stable SPs
possible is the introduction of the cubic-quintic (CQ) nonlinearity, which
includes linear loss (hence the zero background is stable), cubic gain
(provided by a combination of the usual linear gain and saturable
absorption), and additional quintic loss that provides for the overall
stabilization of the model. The CGLE with the CQ nonlinearity was first
proposed (in a 2D form) by Sergeev and Petviashvili \cite{NizhnyNovgorod}. A
stable SP solution in the 1D version of this equation, which is relevant to
modeling fiber lasers, was first reported, in an approximate analytical
form, in \cite{me}. These solutions were found by treating the dissipation
and gain terms in the CGLE as small perturbations added to the usual cubic
NLSE with the anomalous sign of the GVD. Accordingly, the SP was obtained as
a perturbation of the standard NLSE\ soliton. In later works, SPs and their
stability in the CQ CGLE were investigated in a broad region of parameters
\cite{CQCGL1}-\cite{CQCGL5}.

Another possibility to produce stable SPs, which is directly relevant to the
general topic of the present article, is to linearly couple the usual cubic
CGLE to an additional equation which is dominated by the linear loss. A
coupled system of this type was first introduced in Ref. \cite{Herb}, as a
model of a dual-core nonlinear dispersive optical fiber, with linear gain, $%
\gamma _{0}$, in one (\emph{active}) core, and linear loss, $\Gamma _{0}$,
in the other (\emph{passive}) one:
\begin{eqnarray}
iu_{z}+\frac{1}{2}u_{\tau \tau }+|u|^{2}u-i\gamma _{0}u-i\gamma _{1}u_{\tau
\tau }+\kappa v &=&0,  \label{uIntro} \\
iv_{z}+(1/2)v_{\tau \tau }+|v|^{2}v+i\Gamma _{0}v+\kappa u &=&0.
\label{vIntro}
\end{eqnarray}%
Here, $u$ and $v$ are, respectively, envelopes of the electromagnetic waves
in the active and passive cores, $z$ and $\tau $ are, as above, the
propagation distance and reduced time, cf. Eqs. (\ref{ucoupler}) and (\ref%
{vcoupler}), $\kappa $ is the constant of the inter-core coupling, and $%
\gamma _{1}$ accounts for dispersive loss in the active core (in other
words, $\gamma _{1}$ determines the \emph{bandwidth-limited} character of
the linear gain). The model assumes the usual self-focusing Kerr
nonlinearity and anomalous GVD in the fiber, with the respective
coefficients scaled to be $1$. The gain in the dual-core fiber can be
experimentally realized, similar to the usual fiber lasers, by means of
externally pumped resonant dopants \cite{doping-coupler}. Actually, the
dual-core fiber may be fabricated as a symmetric one, with both cores doped,
while only one core is pumped by an external light source, which gives rise
to the gain in that core.

A more general system of linearly-coupled CGLEs applies to a system of
parallel-coupled plasmonic waveguides \cite{Dima}. The system was further
extended for coupled 2D CGLEs, representing stabilized laser cavities \cite%
{Pavel,vort1,vort2}.

It was first theoretically predicted in Refs. \cite{Herb1} and \cite{Herb2}
that, adding the parallel-coupled passive core (with loose ends) to the
soliton-generating fiber laser, one can improve the stability of the output:
while the soliton, being a self-trapped nonlinear mode, remains essentially
confined to the active core, small-amplitude noise easily couples to the
passive one, where it is radiated away through the loose ends.
Independently, a similar dual-core system, with loss in the additional core
but without gain in the main one, was proposed as an optical filter cleaning
solitons from noise \cite{Aussie}. The basic idea is that, due to the action
of the self-focusing nonlinearity, the soliton keeps itself in the core in
which it is propagating, while the linear noise tunnels into the parallel
core, where it is suppressed by the loss. It was found that the best
efficiency of the filtering is attained not with very strong loss ($\Gamma
_{0}$) in the extra core, but rather at $\Gamma _{0}\sim \kappa $, see Eq. (%
\ref{vIntro}) \cite{Aussie}. Another possible application of the dual-core
system with the gain in the \emph{straight core} (the one into which the
input signal is coupled) and losses in the \emph{cross core} (the one
linearly coupled to the straight core)\ was proposed for the design of a
nonlinear amplifier of optical signals: a very weak (linear) input would
pass into the cross core, and would be lost there, while the input whose
power exceeds a certain threshold, making it a sufficiently nonlinear mode,
stays in the straight core, being amplified there \cite{amplifier}.

In Ref. \cite{Herb}, the possibility of the existence of stable SPs in the
system of Eqs. (\ref{uIntro}) and (\ref{vIntro}) was predicted in the
framework of an analytical approximation, that treated both the coupling and
gain/loss terms as small perturbations. The stability of the so predicted
pulses was then verified by direct simulations \cite{Javid1}. Further, it
was demonstrated that the system may be simplified, by dropping the
nonlinear and GVD terms in Eq. (\ref{vIntro}), where they are insignificant
(the nonlinearity may be omitted as the amplitude of the component in the
passive core is small, and the GVD is negligible, as the linear properties
of the passive core are dominated by the loss term). On the other hand, an
extra linear term, namely, a phase-velocity mismatch between the cores,
should be added to Eq. (\ref{vIntro}), as the respective effect may be
essential (see below). As a result, a pair of SP solutions for the
simplified system was found in an \emph{exact analytical form} (which is
displayed below),\emph{\ }the pulse with a larger amplitude being stable in
a vast parameter region, while its counterpart with the smaller amplitude is
always unstable \cite{Javid-exact,Athens}. The basic results are presented,
in some detail, in subsections following below. A detailed review of these
and related results can be found in Ref. \cite{Chaos-review}.

\subsection{The exact SP\ (solitary-pulse) solution}

The most fundamental coupled system, which has the cubic nonlinearity in the
active core only, is based on the following system, which is simplified in
comparison with original equations (\ref{uIntro}) and (\ref{vIntro}) \cite%
{Javid-exact}:%
\begin{equation}
iu_{z}+\left( \frac{1}{2}-i\gamma _{1}\right) u_{\tau \tau }+\left( \sigma
+i\gamma _{2}\right) |u|^{2}u-i\gamma _{0}u+v=0,  \label{u}
\end{equation}%
\begin{equation}
iv_{z}+k_{0}v+i\Gamma _{0}v+u=0,  \label{v}
\end{equation}%
where $\kappa =1$ is fixed by means of scaling, $\sigma =+1$ and $-1$
correspond to the anomalous and normal GVD in the active core (assuming that
the actual nonlinearity is self-focusing, which is the case in optical
fibers, this sign parameter may be placed in front of the cubic term, as
written in Eq. (\ref{u}), although $\sigma $ originally appears in front of
the second derivative), $\gamma _{2}\geq 0$ accounts for cubic loss
(two-photon absorption), and $k_{0}$ is the above-mentioned phase-velocity
mismatch between the cores.

The \emph{exact} SP solution to Eqs. (\ref{u}) and (\ref{v}) can be found in
the analytical form suggested by the well-known solution \cite%
{Hocking,Lennart} of the cubic CGLE:
\begin{equation}
\left\{ u,v\right\} =\left\{ A,B\right\} e^{ikz}~\left[ \mathrm{sech}\left(
\chi \tau \right) \right] ^{1+i\mu },  \label{exact}
\end{equation}%
where all the constants but $B$ are real. Coefficient $\mu $, which
determines the \emph{chirp} of the pulse, is
\begin{equation}
\mu =\frac{\sigma \sqrt{9\left( 1-2\sigma \gamma _{1}\gamma _{2}\right)
^{2}+8\left( 2\gamma _{1}+\sigma \gamma _{2}\right) ^{2}}-3\left( 1-2\sigma
\gamma _{1}\gamma _{2}\right) }{2\left( 2\gamma _{1}+\sigma \gamma
_{2}\right) }.  \label{mu}
\end{equation}%
The complex and real amplitudes, $B$ and $A$, are given by expressions
\begin{gather}
B=\left( k-k_{0}-i\Gamma _{0}\right) ^{-1}A,  \label{B} \\
A^{2}=\sigma \frac{\left[ \left( 1-2\sigma \gamma _{1}\gamma _{2}\right)
\left( 2-\mu ^{2}\right) +3\mu \left( 2\gamma _{1}+\sigma \gamma _{2}\right) %
\right] \chi ^{2}}{2\left( 1+2\gamma _{2}^{2}\right) },  \label{A^2}
\end{gather}%
with the two remaining real parameters $\chi $ and $k$ determined by one
complex equation,
\begin{equation}
k+i\gamma _{0}-\left( k-k_{0}-i\Gamma _{0}\right) ^{-1}=\left( \frac{1}{2}%
-i\gamma _{1}\right) \left( 1+i\mu \right) ^{2}\chi ^{2}.  \label{eq}
\end{equation}

In the further analysis of the SP solutions, one may set $\gamma _{2}=0$, as
the two-photon absorption is insignificant in silica fibers. Then, $\chi
^{2} $ can be eliminated from Eq. (\ref{eq}),
\begin{equation}
\chi ^{2}=\frac{8\gamma _{0}\gamma _{1}}{8\gamma _{1}^{2}+3-\sigma \sqrt{%
9+32\gamma _{1}^{2}}}\left( 1-\frac{\Gamma _{0}}{\gamma _{0}\left[ \left(
k-k_{0}\right) ^{2}+\Gamma _{0}^{2}\right] }\right) ,  \label{chi^2}
\end{equation}%
and one arrives at a final cubic equation for $k$:%
\begin{equation}
k\left[ \left( k-k_{0}\right) ^{2}+\Gamma _{0}^{2}\right] -\left(
k-k_{0}\right) =\frac{\sigma \sqrt{9+32\gamma _{1}^{2}}}{2\gamma _{1}}\left[
\gamma _{0}\left( k-k_{0}\right) ^{2}-\Gamma _{0}\left( 1-\gamma _{0}\Gamma
_{0}\right) \right] ,  \label{k}
\end{equation}%
which may give rise to one or three real solutions. Physical solutions are
those which make expression (\ref{chi^2}) positive. In particular, the
number of the physical solutions changes when expression (\ref{chi^2})
vanishes, which happens at $k_{0}^{2}=\left( \gamma _{0}\Gamma _{0}\right)
^{-1}\left( 1-\gamma _{0}\Gamma _{0}\right) \left( \Gamma _{0}-\gamma
_{0}\right) ^{2}$.

The above results may be cast in a more explicit form in the case of no
wavenumber mismatch between the cores, $k_{0}=0$. Note that the SP solution
is of interest if it is stable, a necessary condition for which is the
stability of the zero background, i.e., the trivial solution, $u=v=0$. If $%
k_{0}=0$, necessary conditions for the stability of the zero solution are $%
\gamma _{0}<\Gamma _{0}$ and $\gamma _{0}\Gamma _{0}<1$. It is natural to
focus on the case when gain $\gamma _{0}$ is close to the maximum value, $%
\left( \gamma _{0}\right) _{\max }\equiv 1/\Gamma _{0}$, admitted by the
latter condition, i.e.,%
\begin{equation}
0<1-\gamma _{0}\Gamma _{0}\ll 1  \label{gG}
\end{equation}%
(then, condition $\gamma _{0}<\Gamma _{0}$ reduces to $\Gamma _{0}>1$). In
this case, Eq. (\ref{k}) may be easily solved. The first root has small $k$,
which yields unphysical solutions, with $\chi ^{2}<0$. Two other roots for $%
k $ are physically relevant ones, in which $\gamma _{0}$ may be replaced by $%
1/\Gamma _{0}$, due to relation (\ref{gG}):
\begin{eqnarray}
4k &=&\frac{\sigma \sqrt{9+32\gamma _{1}^{2}}}{\Gamma _{0}\gamma _{1}}\pm
\sqrt{\frac{9+32\gamma _{1}^{2}}{\Gamma _{0}^{2}\gamma _{1}^{2}}-16\left(
\Gamma _{0}^{2}-1\right) },  \label{newsolk} \\
\chi ^{2} &=&\frac{8\gamma _{1}k^{2}\left( k^{2}+\Gamma _{0}^{2}\right) ^{-1}%
}{\Gamma _{0}\left( 3+8\gamma _{1}^{2}-\sigma \sqrt{9+32\gamma _{1}^{2}}%
\right) }.  \label{newsolchi}
\end{eqnarray}%
Expression (\ref{newsolchi}) is always positive, while a nontrivial
existence condition for these two solutions follows from Eq. (\ref{newsolk}%
): $k$ must be real, which means that
\begin{equation}
9\gamma _{1}^{-2}+32>16\Gamma _{0}^{2}\left( \Gamma _{0}^{2}-1\right) .
\label{ineq}
\end{equation}%
Thus, Eqs. (\ref{newsolk}) and (\ref{newsolchi}), along with Eqs. (\ref%
{exact}), (\ref{mu}), (\ref{B}), (\ref{A^2}), and (\ref{chi^2}), (\ref{k}),
furnish the SP solutions in the region of the major interest, and Eq. (\ref%
{ineq}) is a fundamental condition which secures the existence of these
solutions.

If condition (\ref{ineq}) holds, one has the following set of solutions: (i)
the stable zero state, (ii) the broader SP with a smaller amplitude,
corresponding to smaller $k^{2}$, i.e., with sign $\pm $ in expression (\ref%
{newsolk}) chosen opposite to $\sigma $, and (iii) the narrower pulse with a
larger amplitude, corresponding to larger $k^{2}$, i.e., with $\pm $ in (\ref%
{newsolk}) chosen to coincide with $\sigma $. Basic principles of the
bifurcation theory \cite{bifurcations} suggest that stable and unstable
solutions alternate, hence, because the trivial solution is stable, the
larger-amplitude narrower pulse ought to be stable too, while the
intermediate broader pulse with the smaller amplitude is always unstable,
playing the role of a \emph{separatrix }between the two \emph{attractors}.
This expectation is, generally, corroborated by numerical results \cite%
{Atai1,Atai2,Javid1,Javid-exact}, as shown in some detail below. Note that
the above-mentioned SP waveform \cite{Hocking,Lennart}, which suggested
ansatz (\ref{exact}) for the exact solutions under the consideration, is, by
itself, \emph{always unstable }as the solution of the single cubic CGL
equation.

\subsection{Special cases of stable SPs (solitary pulses)}

There are two particular cases of physical interest that should be
considered separately. The first corresponds to the model with $\gamma _{1}=0
$ (negligible dispersive loss). In this case, the above SP solution may only
exist in the system with anomalous GVD, $\sigma =+1$. Special consideration
of this case is necessary because the above formulas are singular for $%
\gamma _{1}=0$. An explicit result for this situation can be obtained \emph{%
without} adopting condition (\ref{gG}): the solutions take the form of Eq. (%
\ref{exact}) with $\mu =0$ (no chirp), Eq. (\ref{A^2}) being replaced by $%
A^{2}=\chi ^{2}$, while Eqs. (\ref{eq}) and (\ref{A^2}) become%
\begin{equation*}
k-k_{0}=\pm \sqrt{\Gamma _{0}\gamma _{0}^{-1}\left( 1-\gamma _{0}\Gamma
_{0}\right) },\,
\end{equation*}%
\begin{equation}
\chi ^{2}=2k_{0}\pm 2\left( \gamma _{0}+\Gamma _{0}\right) \sqrt{\left(
\gamma _{0}\Gamma _{0}\right) ^{-1}\left( 1-\gamma _{0}\Gamma _{0}\right) }.
\label{zerok}
\end{equation}%
Two solutions corresponding to both signs in Eqs. (\ref{zerok}) exist
simultaneously, i.e., $\chi ^{2}>0$ holds for both of them, provided that $%
k_{0}^{2}>\left( \gamma _{0}\Gamma _{0}\right) ^{-1}\left( 1-\gamma
_{0}\Gamma _{0}\right) \left( \gamma _{0}+\Gamma _{0}\right) ^{2}$. However,
one can check that this inequality contradicts the stability conditions of
the zero state, therefore only \emph{one} solution given by Eqs. (\ref{zerok}%
) may exist in the case of interest. General principles of the bifurcation
theory \cite{bifurcations} suggest that this single nontrivial solution is
automatically unstable in the case of $\gamma _{1}=0$, once the trivial one
is stable.

The other specially interesting case is that of zero GVD, corresponding to
the physically important situation when the carrier wavelength is close to
the zero-dispersion point \cite{Agrawal} of the optical fiber. In this case,
Eq. (\ref{v}) does not change its form, while Eq. (\ref{u}) is replaced by
\begin{equation}
iu_{z}-iu_{\tau \tau }+|u|^{2}u-i\gamma _{0}u+v=0  \label{zerodisp}
\end{equation}%
(in the absence of the GVD, both signs of $\sigma $ are equivalent, hence $%
\sigma =+1$ is fixed here, and normalization $\gamma _{1}\equiv 1$ may be
adopted). Explicit solutions can be obtained, as well as in the general case
considered above, by assuming $k_{0}=0$ and taking $\gamma _{0}\Gamma _{0}$
close to $1$. Then, expressions (\ref{A^2}) and (\ref{mu}) are replaced by $%
\mu =\sqrt{{2}}$, $A^{2}=3\sqrt{2}\chi ^{2}$, while solutions (\ref{newsolk}%
) and (\ref{newsolchi}) take the form of
\begin{equation}
k=\sqrt{2}\Gamma _{0}^{-1}\pm \sqrt{2\Gamma _{0}^{-2}-\Gamma _{0}^{2}+1}%
,\;\,\chi ^{2}=\Gamma _{0}^{-1}\left( \Gamma _{0}^{2}+k^{2}\right)
^{-1}k^{2},  \label{zerodispchi}
\end{equation}%
and the existence condition (\ref{ineq}) becomes very simple, $\Gamma
_{0}^{2}<2$. Thus, in the zero-GVD case both SP solutions (\ref{zerodispchi}%
) exist simultaneously, suggesting that the one with the larger value of $%
k^{2}$ may be \emph{stable}.

Lastly, it may be interesting to consider another particular case, with
normal GVD, $\sigma =-1$, and small dispersive-loss coefficient, $\gamma
_{1}\ll 1$. In this case, there are two nontrivial SP solutions, the one
with the larger amplitude, that has the chance to be stable, being%
\begin{equation}
\mu =-\frac{3}{2}\gamma _{1}^{-1},A^{2}=\frac{3}{2}\left( \gamma _{1}\Gamma
_{0}\right) ^{-1},k=-\frac{3}{2}\gamma _{1}^{-1},\chi ^{2}=\frac{4}{3}\left(
\Gamma _{0}^{-1}\gamma _{1}\right) .  \label{normalchi}
\end{equation}%
The large value of $\mu $ in this solution implies that the pulse is
strongly chirped. Obviously, the latter solution disappears in the limit of $%
\gamma _{1}\rightarrow 0$, in accordance with the above-mentioned negative
result (no stable SP) for the case of $\gamma _{1}=0$.

\subsection{Stability of the solitary pulses and dynamical effects}

As mentioned above, the SP~cannot be stable unless its background, $u=v=0$,
is stable. To explore the stability of the zero solution, infinitesimal
perturbations are substituted in the linearized version of Eqs.~(\ref{u})
and (\ref{v}): $\left\{ u,v\right\} =\left\{ u_{1},v_{1}\right\}
e^{i(qz-\omega t)}$, where $\omega $ and $q$ are an arbitrary real frequency
and the corresponding propagation constant (generally speaking, a complex
one) . The stability condition is $\mathrm{Im\,}(q)=0$, which must hold at
all real $\omega $. This condition leads an inequality that should be valid
at all $\omega ^{2}\geq 0$,
\begin{equation}
\Gamma _{0}(\gamma _{0}-\gamma _{1}\omega ^{2})\left[ 1+\frac{\left(
2k_{0}+\omega ^{2}\right) ^{2}}{4\left( \Gamma _{0}-\gamma _{0}+\gamma
_{1}\omega ^{2}\right) ^{2}}\right] \leq 1.  \label{stab1}
\end{equation}%
If the GVD coefficient is zero, i.e., Eq. (\ref{u}) is replaced by Eq. (\ref%
{zerodisp}), expression $\left( 2k_{0}+\omega ^{2}\right) ^{2}$ in Eq. (\ref%
{stab1}) is replaced by $4k_{0}^{2}$. An obvious corollary of Eq. (\ref%
{stab1}) is
\begin{equation}
\gamma _{0}<\Gamma _{0},  \label{gg}
\end{equation}%
i.e., the trivial solution may be stable only if the loss in the passive
core is stronger than the gain in the active one. Further, in the case of
zero wavenumber mismatch between the cores, $k_{0}=0$, which was considered
above (see Eqs. (\ref{newsolk}) and (\ref{newsolchi})), a simple necessary
stability condition is obtained from Eq. (\ref{stab1}) at $\omega =0$ (as
was already mentioned above, see Eq. (\ref{gG})): $\gamma _{0}\Gamma _{0}>1$.

Stability regions of the SP solutions in the parameter space of the system
were identified in Refs. \cite{Javid1} and \cite{Athens} in a numerical
form, combining the analysis of the necessary stability condition of the
zero background, given by Eq. (\ref{stab1}), and direct simulations of Eqs. (%
\ref{u}) and (\ref{v}) for perturbed SPs. As said above, the case of major
interest is the one with $k_{0}=\gamma _{2}=0$ and $\gamma _{0}\Gamma _{0}$
close to $1$. For this case, the stability regions are displayed in Fig. \ref%
{fig8}, and separately in Fig. \ref{fig9} for the zero-GVD system. In each
case, there is a single stable SP (but the system is a \emph{bistable} one,
as the stable SP always coexists with the stable zero solution).
\begin{figure}[tbp]
\begin{center}
\includegraphics[height=10cm]{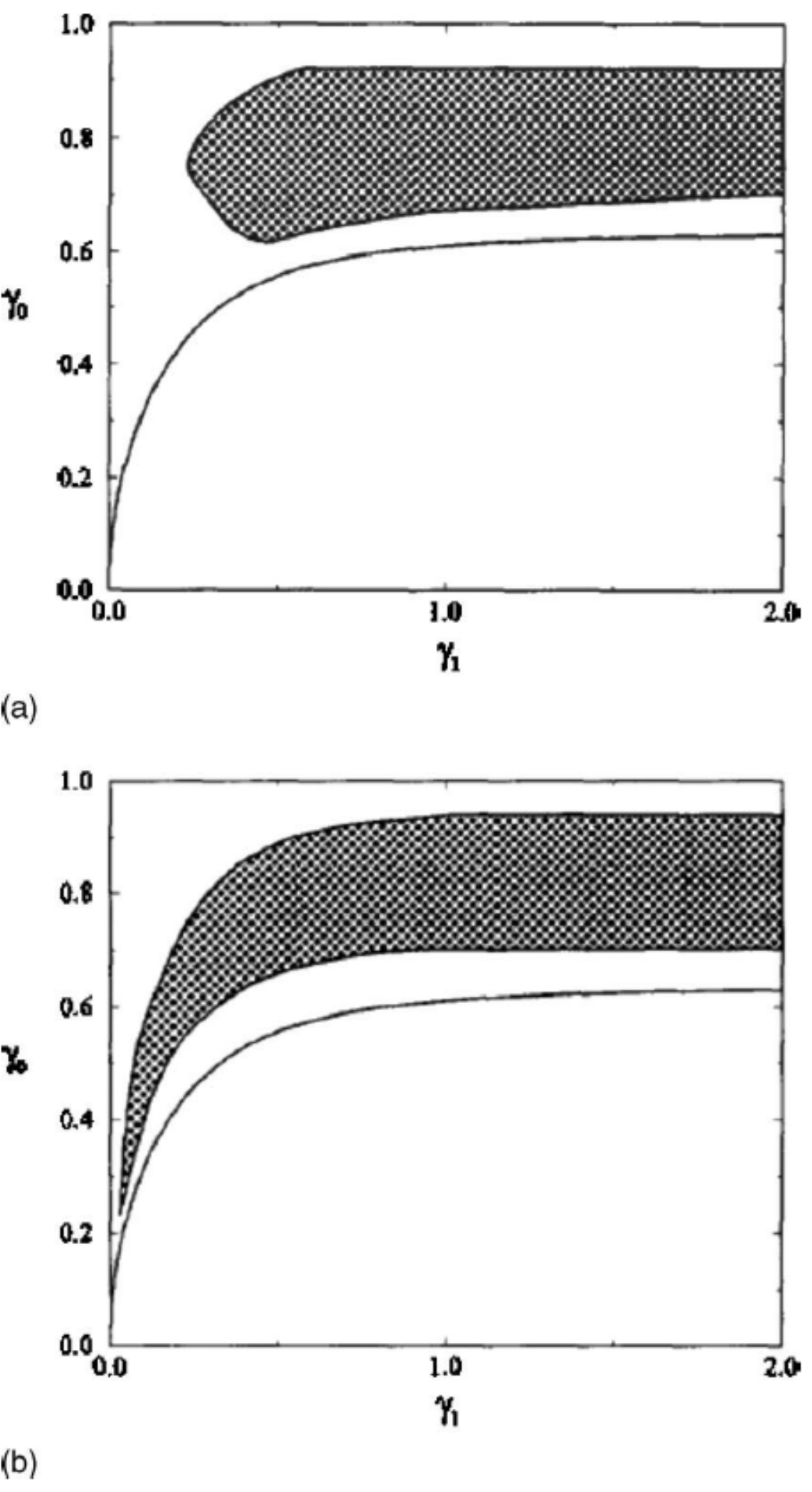}
\end{center}
\caption{The stability region (shaded) in the parameter plane of the exact
solitary-pulse solution (\protect\ref{exact}) of Eqs. (\protect\ref{u}) and (%
\protect\ref{v}), for the case of $k_{0}=\protect\gamma _{2}=0$ and $\protect%
\gamma _{0}\Gamma _{0}=0.9$, as per Refs. \protect\cite{Javid-exact} and
\protect\cite{Chaos-review}. Panels (a) and (b) display results for the
anomalous and normal GVD, respectively, i.e., $\protect\sigma =+1$ and $%
\protect\sigma =-1$ in Eq. (\protect\ref{u}). The separate curve shows the
existence boundary given by Eq. (\protect\ref{ineq}).}
\label{fig8}
\end{figure}
\begin{figure}[tbp]
\begin{center}
\includegraphics[height=10cm]{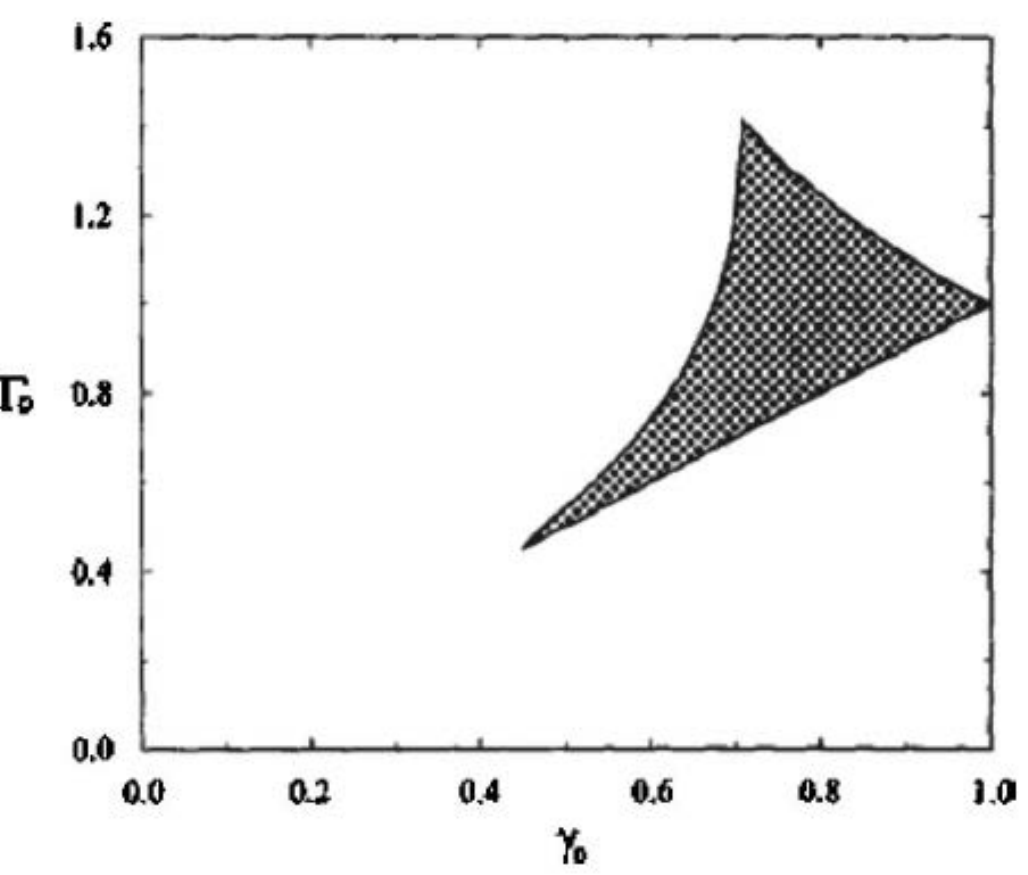}
\end{center}
\caption{The same as in Fig. \protect\ref{fig8}, but for the exact
solitary-pulse solution (\protect\ref{zerodispchi}), in the system with zero
dispersion (as per Refs. \protect\cite{Javid-exact} and \protect\cite%
{Chaos-review}).}
\label{fig9}
\end{figure}

\begin{figure}[tbp]
\begin{center}
\includegraphics[height=10cm]{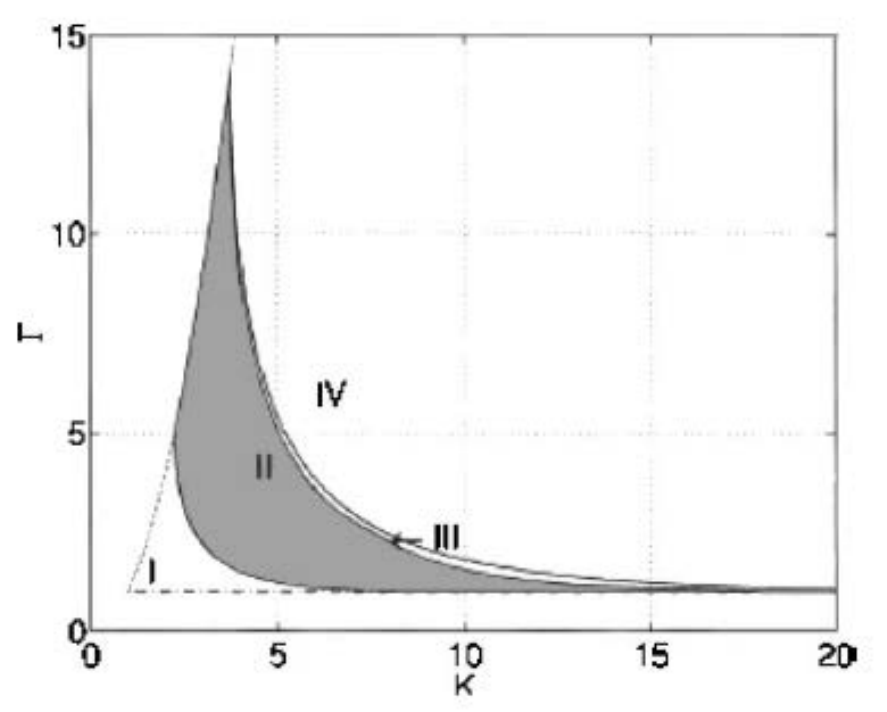}
\end{center}
\caption{Stability regions of the zero solution, and existence and stability
regions for the exact solitary-pulse solution (\protect\ref{exact}) of Eqs. (%
\protect\ref{u}) and (\protect\ref{v}), as per Refs. \protect\cite%
{Javid-exact} and \protect\cite{Chaos-review}, in the plane of parameters $%
K\equiv 1/\protect\gamma _{0}$ and $\Gamma \equiv $ $\Gamma _{0}/\protect%
\gamma _{0}$, in the model with the normal GVD$\ (\protect\sigma =-1$), $%
k_{0}=$ $\protect\gamma _{2}=0$, and $\protect\gamma _{1}=1/18$. Region~I:
the zero background is unstable. Region~II: the solitary pulse is stable.
Region~III: the zero background is stable, while the solitary pulse is not,
decaying to zero in direct simulations. Region~IV: the solitary-pulse
solution does not exist. In the region located outside of region I, which is
bordered by curves $\protect\gamma _{0}\Gamma _{0}=1$ (the dotted curve) and
$\Gamma _{0}=\protect\gamma _{0}$ (the dotted-dashed curve), the zero
solution is certainly unstable, as condition~(\protect\ref{gG}) does not
hold in that region. In region I, zero background is unstable even though
condition~(\protect\ref{gG}) holds in this region. }
\label{fig10}
\end{figure}

Another cross section of the stability region in the full three-dimensional
parameter space of the model is represented by region II in Fig. \ref{fig10}%
, for the normal-GVD case ($\sigma =-1$), with $\gamma _{1}=1/18$ (this
value is a typical one corresponding to physically relevant parameters of
optical fibers). Note that this plot reveals a very narrow region (area
III), in which the zero solution is stable, while the SP\ is not.

\subsection{Interactions between solitary pulses}

\begin{figure}[tbp]
\begin{center}
\includegraphics[height=11.5cm]{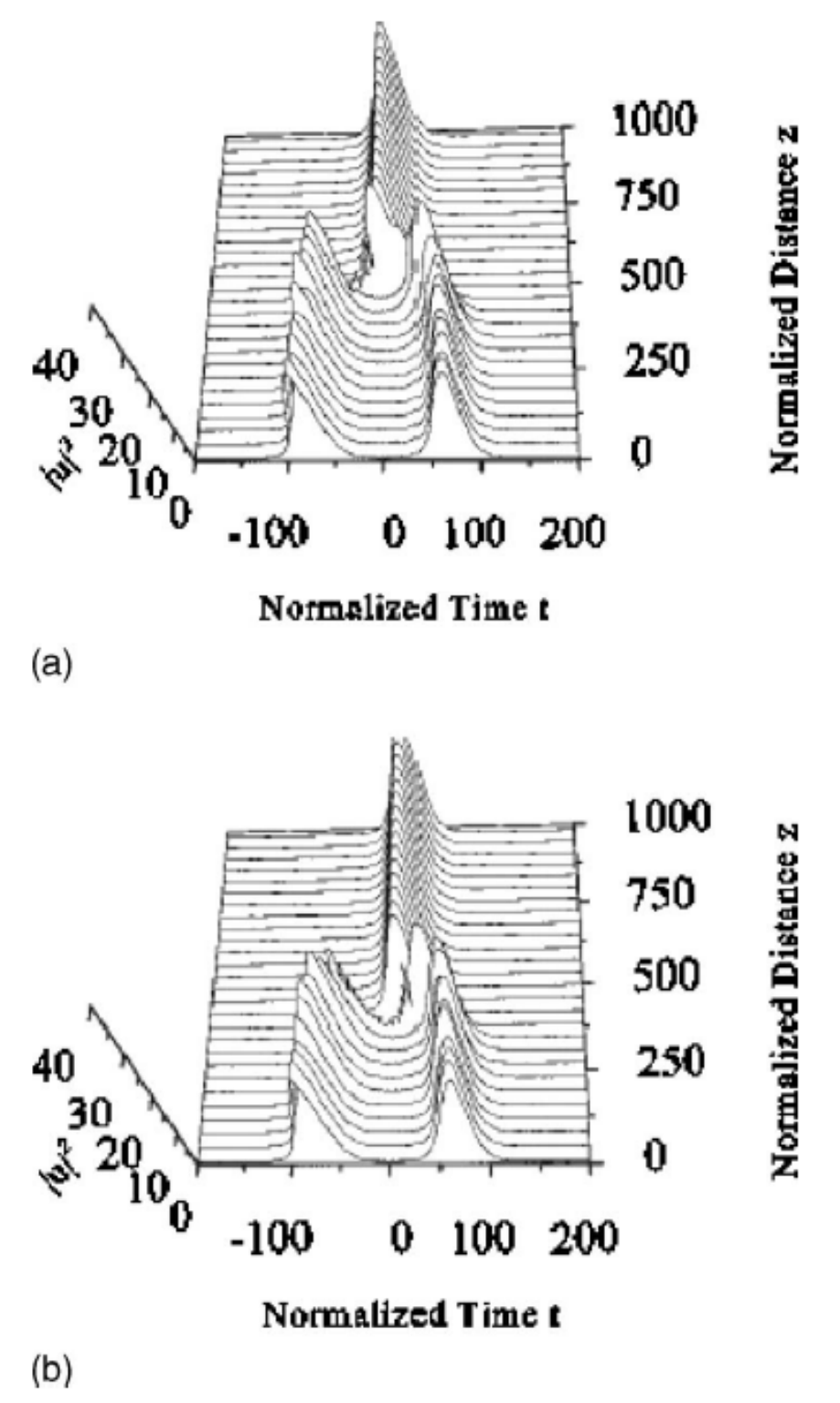}
\end{center}
\caption{Merger of chirped solitary pulses in the system of Eqs. (\protect
\ref{u}) and (\protect\ref{v}) with normal GVD ($\protect\sigma =-1$). Other
parameters are $k_{0}=\protect\gamma _{2}=0$, $\protect\gamma _{0}=0.2$, $%
\Gamma _{0}=0.8$, and $\protect\gamma _{1}=1/18$. The initial phase shift
between the pulses is $\Delta \protect\phi =0$ (a) and $\Delta \protect\phi =%
\protect\pi $ (b). The results are displayed as per Refs. \protect\cite%
{JavidBoundStates} and \protect\cite{Chaos-review}.}
\label{fig11}
\end{figure}

It is well known that the sign of the interaction between ordinary solitons
is determined by the phase shift between them, $\Delta \phi $: the
interaction is attractive for in-phase soliton pairs, with $\Delta \phi =0$,
and repulsive for $\Delta \phi =\pi $ \cite{RMP}. \ However, this rule does
not apply to the SPs in the present model with the normal sign of the GVD ($%
\sigma =-1$ in Eq. (\ref{u})), which feature strong chirp in their phase
structure (see, e.g., the expression for chirp $\mu $ in solution (\ref%
{normalchi}), with $\gamma _{1}\ll 1$). It was found \cite{Athens} that,
irrespective of the value of $\Delta \phi $, the SPs in the normal-GVD model
\emph{attract} each other, and eventually merge into a single pulse, as
shown in Fig. \ref{fig11}.

On the other hand, in-phase pairs of the SPs in the anomalous-GVD system,
with $\sigma =+1$, readily form robust bound states \cite{JavidBoundStates}.
Three-pulse bound states were found too, but they are unstable against
symmetry-breaking perturbations, which split them according to the scheme $%
3\rightarrow 2+1$.

\subsection{CW (continuous-wave)\ states and dark solitons
(\textquotedblleft holes")}

In addition to the SPs, Eqs. (\ref{u}) and (\ref{v}) also admit CW states
with constant amplitudes,
\begin{equation}
u=a\,\exp \left( ikz-i\omega \tau \right) ,\;v=b\,\exp \left( ikz-i\omega
\tau \right) .  \label{cw}
\end{equation}%
The propagation constant and amplitudes of this solution can be easily found
in the case of $\gamma _{2}=0$:
\begin{equation*}
b=\left( k-k_{0}-i\Gamma _{0}\right) ^{-1}a,
\end{equation*}%
\begin{equation*}
k-k_{0}=\pm \sqrt{\Gamma _{0}\tilde{\gamma}_{0}^{-1}\left( 1-\Gamma _{0}%
\tilde{\gamma}_{0}\right) },\,
\end{equation*}%
\begin{equation}
\sigma a^{2}=k_{0}\pm \sqrt{\left( \Gamma _{0}\tilde{\gamma}_{0}\right)
^{-1}\left( 1-\Gamma _{0}\tilde{\gamma}_{0}\right) }\left( \Gamma _{0}-%
\tilde{\gamma}_{0}\right) ,  \label{solCW}
\end{equation}%
where $\tilde{\gamma}_{0}\equiv $ $\gamma _{0}-\gamma _{1}\omega ^{2}$.
According to this, at given $\omega $ there may exist two CW states with
different amplitudes, provided that $k_{0}^{2}\geq \left( \Gamma _{0}\tilde{%
\gamma}_{0}\right) ^{-1}\left( 1-\Gamma _{0}\tilde{\gamma}_{0}\right) \left(
\Gamma _{0}-\tilde{\gamma}_{0}\right) ^{2}$, and a single one in the
opposite case. The CWs may be subject to the MI (modulational instability),
which was investigated in detail \cite{dark,Ponz}. In particular, all CWs
are unstable at $k_{0}=0$, although the character of the MI is different for
the normal and anomalous signs of the GVD \cite{Ponz}. Direct simulations
demonstrate that the development of the MI splits the CW state into an array
of SPs.

\begin{figure}[tbp]
\includegraphics[width=4in]{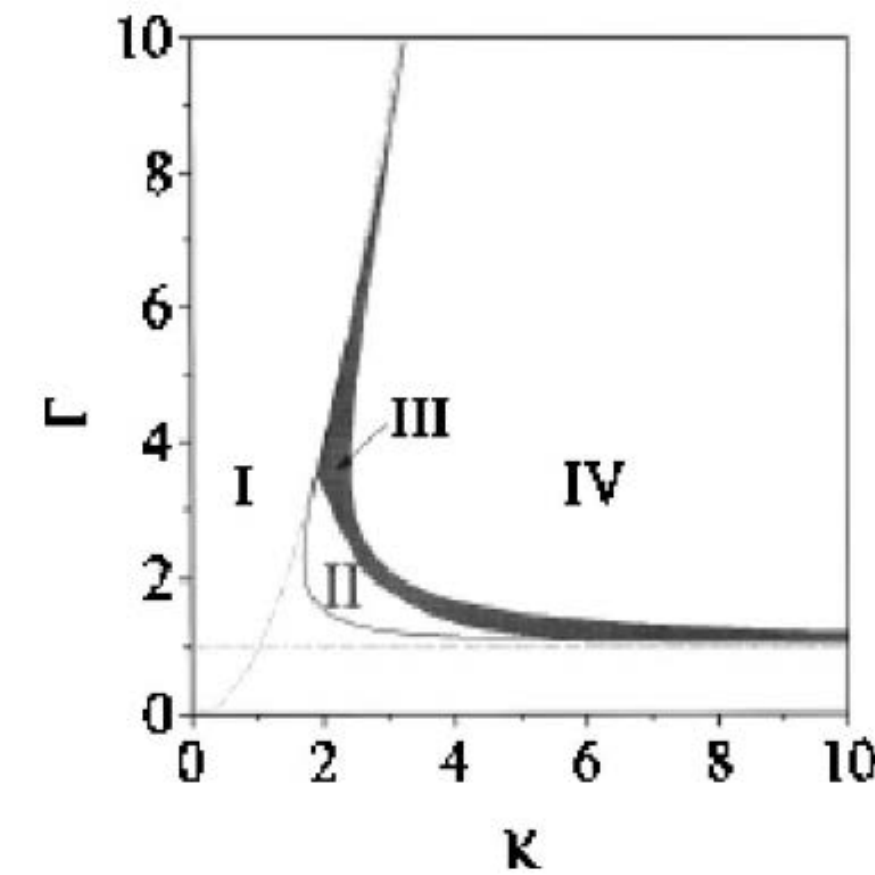}
\caption{Regions of the existence and stability of the CW state and dark
soliton (\protect\ref{hole}), produced by Eqs. (\protect\ref{u}) and (%
\protect\ref{v}) in the same parameter plane as in Fig. \protect\ref{fig10},
in the case of $\protect\sigma =-1$, $\protect\gamma _{2}=0$, $\protect%
\gamma _{1}=1/7$, and $k_{0}=2\protect\gamma _{2}$. The existence region of
the CW solution is confined by the dashed lines, $\Gamma _{0}=1$ and $%
\protect\gamma _{0}\Gamma _{0}=1$. In regions I and IV, the CW is unstable.
In region II, it is stable, while the dark soliton is not. In region III,
both the CW and dark-soliton solutions are stable. The results are displayed
as per Refs. \protect\cite{Athens} and \protect\cite{Chaos-review}.}
\label{fig12}
\end{figure}

At $k_{0}\neq 0$, there is a parameter region in the normal-GVD regime (with
$\sigma =-1$) where the CW solutions are stable, which suggests to explore
solutions in the form of dark solitons (which are frequently called
\textquotedblleft holes", in the context of the CGLEs \cite{holes}). Such
solutions of Eqs. (\ref{u}) and (\ref{v}) can be found in an exact
analytical form based on the following ansatz \cite{dark-CGL} (cf. the form
of exact solution (\ref{exact}) for the bright SP):%
\begin{equation}
u=\frac{\left( 2-e^{2\chi \tau }\right) A}{\left( 1+e^{2\chi \tau }\right)
^{1-i\mu }}e^{ikz-i\mu \chi \tau },v=\frac{u}{k_{0}-k+i\Gamma _{0}},
\label{hole}
\end{equation}%
with $\mu =\left( 3/4\gamma _{1}\right) +\sqrt{\left( 9/4\gamma _{1}\right)
^{2}+2}$ (in the case of $\gamma _{2}=0$), the other parameters, $A,\chi $,
and $k$, being determined by cumbersome algebraic equations. Numerical
analysis demonstrates that dark solitons (\ref{hole}) are stable in a small
parameter region, as shown in Fig. \ref{fig12}.

\subsection{Evolution of solitary pulses beyond the onset of instability}

\begin{figure}[tbp]
\includegraphics[width=5in]{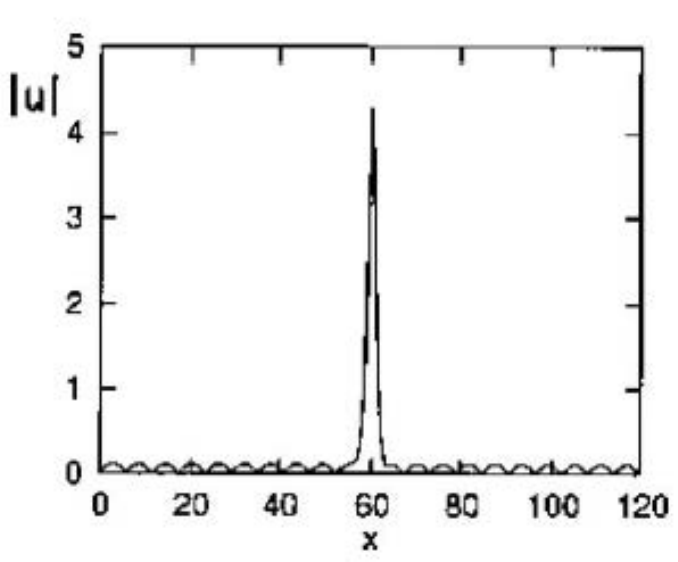}
\caption{An example of a stable solitary pulse in the system of Eqs. (%
\protect\ref{u}) and (\protect\ref{v}), with the anomalous GVD ($\protect%
\sigma =+1$), which exists on top of the small-amplitude background
featuring regular oscillations, in the case when the zero background is
weakly unstable (the respective overcriticality is $\protect\epsilon =0.025$%
, see Eq. (\protect\ref{eps})). Parameters are $k_{0}=0$, $\protect\gamma %
_{0}=0.54$, $\Gamma _{0}=1.35$, $\protect\gamma _{1}=0.18$. The results are
displayed as per Refs. \protect\cite{Hidetsugu} and \protect\cite%
{Chaos-review}.}
\label{fig13}
\end{figure}

Direct simulations of the system with normal GVD, $\sigma =-1$ in Eq. (\ref%
{v}), demonstrate that unstable SPs (in the case when stable solutions do
not exist) either decay to zero (if the zero background is stable), or blow
up, initiating a transition to a \textquotedblleft turbulent" state, if the
background is unstable. A different behavior of unstable SPs was found in
Ref. \cite{Hidetsugu} in the model with the anomalous GVD ($\sigma =+1$). If
the instability of the zero background is weak, it does not necessarily lead
to the blow-up. Instead, it may generate a small-amplitude background field
featuring regular oscillations. In that case, the SP sitting on top of such
a small-amplitude background remains completely stable, as shown in Fig. \ref%
{fig13}. The proximity of this state to the stability border is
characterized by the \emph{overcriticality parameter},
\begin{equation}
\epsilon \equiv \left( \gamma _{0}-\left( \gamma _{0}\right) _{\mathrm{cr}%
}\right) /\left( \gamma _{0}\right) _{\mathrm{cr}},  \label{eps}
\end{equation}%
where $\left( \gamma _{0}\right) _{\mathrm{cr}}$is the critical size of the
linear gain at which the instability of the zero solution sets in, at given
values of $\gamma _{1}$,~$k_{0}$, and $\Gamma _{0}$. An example of the
stable pulse found on top of the finite background, which is displayed in
Fig. \ref{fig13}, pertains to $\epsilon =0.025$.

\begin{figure}[tbp]
\includegraphics[width=6in]{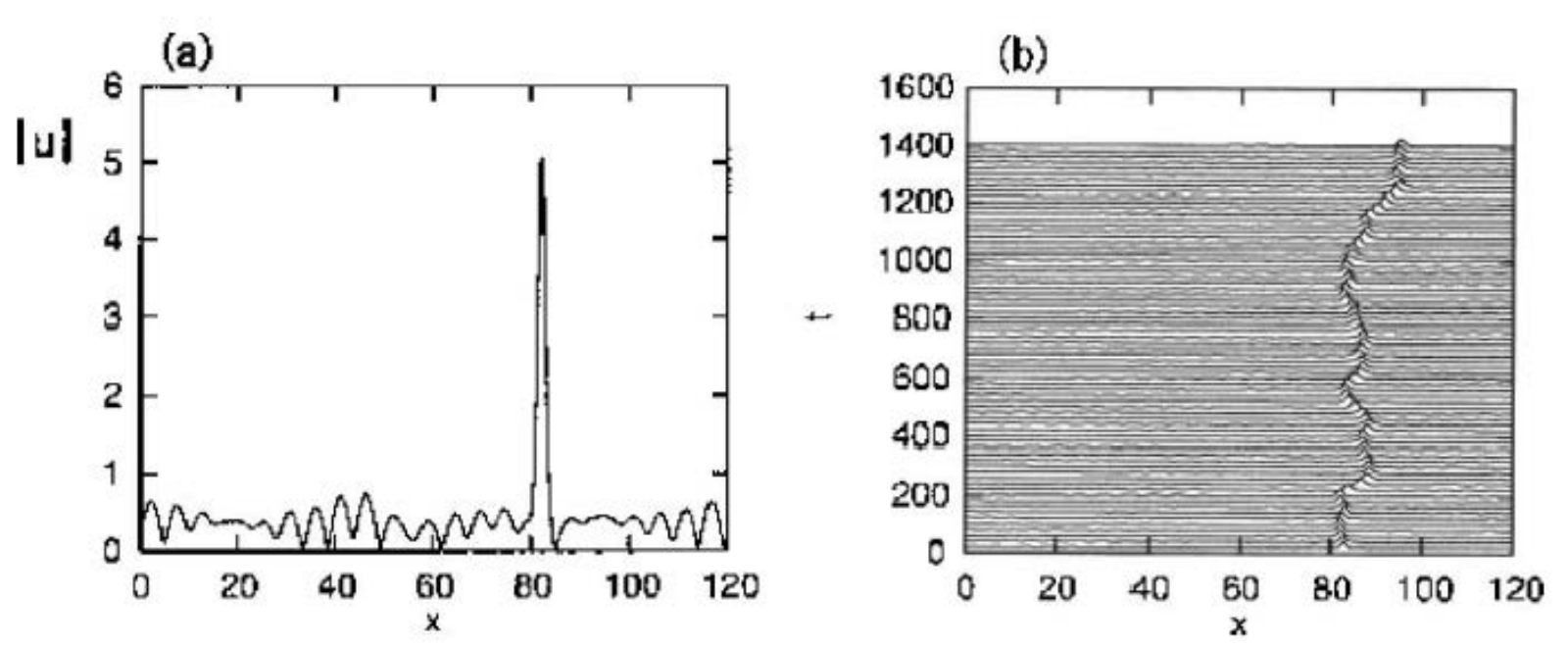}
\caption{(a) An example of a solitary pulse which remains stable, as a
whole, on top of the background featuring chaotic oscillations, at
overcriticality $\protect\epsilon =0.157$. Parameters are the same as in
Fig. \protect\ref{fig13}, except for $\protect\gamma _{0}=0.61.$ (b) The
random walk of the stable pulse from (a), driven by its interaction with the
chaotic background. The results are displayed as per Refs. \protect\cite%
{Hidetsugu} and \protect\cite{Chaos-review}.}
\label{fig14}
\end{figure}

At larger but still moderate values of the overcriticality, such as $%
\epsilon =0.157$ in Fig. \ref{fig14}, the background oscillations become
chaotic, while keeping a relatively small amplitude. As a result of the
interaction with this chaotic background, the SP remains stable as a whole,
featuring a random walk. The walk shows a typically diffusive behavior, with
the average squared shift in the $\tau $-direction growing linearly with $z$
\cite{Hidetsugu}. The randomly walking pulses may easily form bound states,
which then feature synchronized random motion. Finally, at essentially
larger values of the overcriticality, $\epsilon \gtrsim 1$, the system goes
over into a turbulent state, which may be interpreted as a chaotic gas of
solitary pulses \cite{Hidetsugu}.

\section{Soliton stability in $\mathcal{PT}$ (parity-time)-symmetric
nonlinear dual-core fibers}

The dual-core fibers with equal (mutually balanced) gain and loss in the
cores offer a natural setting for the realization of the $\mathcal{PT}$
symmetry, in addition to other optical media, where this symmetry was
proposed theoretically \cite{Muga}, \cite{Chr1}, \cite{Chr2}-\cite{Longhi},
\cite{Chr-review}-\cite{review2}, \cite{PT-CQ}, and implemented
experimentally \cite{experiment1,Moti}.

The basic model of the $\mathcal{PT}$-symmetric nonlinear coupler is based
on the equations similar to Eqs. (\ref{ucoupler}) and (\ref{vcoupler}), in
which $\gamma >0$ is the gain-loss coefficient, and the coefficient of the
inter-core coupling ($K$ in Eqs. (\ref{ucoupler}) and (\ref{vcoupler})) is
scaled to be $1$ \cite{PT1,PT2}:%
\begin{eqnarray}
iu_{z}+(1/2)u_{tt}+|u|^{2}u-i\gamma u+v &=&0,  \label{uPT} \\
iv_{z}+(1/2)v_{tt}+|v|^{2}v+i\gamma v+u &=&0.  \label{vPT}
\end{eqnarray}%
Note that the $\mathcal{PT}$-balanced gain and loss in this system
correspond to the border between stable and unstable settings: if the loss
coefficient in Eq. (\ref{vPT}) is replaced by an independent one, $\Gamma >0$
(different from the gain factor $\gamma $ in Eq. (\ref{vPT})), the zero
solution, $u=v=0$, is unstable at $\gamma >\Gamma $, and may be stable at $%
\gamma <\Gamma $, see Eq. (\ref{gg}) above.

Obviously, \emph{any} solution to the NLSE (with a frequency shift), $%
iU_{z}+(1/2)U_{tt}+|U|^{2}U\pm \sqrt{1-\gamma ^{2}}U=0$, gives rise to two
\emph{exact} solutions of the $\mathcal{PT}$-symmetric system, provided that
condition $\gamma \leq 1$ holds:
\begin{equation}
v=\left( i\gamma \pm \sqrt{1-\gamma ^{2}}\right) u=U\left( z,t\right) .
\label{UU}
\end{equation}%
For $\gamma =0$, solutions (\ref{UU}) with $+$ and $-$ amount, respectively,
to symmetric and antisymmetric modes in the dual-core coupler, therefore the
respective solutions (\ref{UU}) may be called $\mathcal{PT}$-symmetric and $%
\mathcal{PT}$-antisymmetric ones. In the limit of $\gamma =1$, two solutions
(\ref{UU}) reduce to a single one, $v=iu=U(z,t)$. In particular, $\mathcal{PT%
}$-symmetric and antisymmetric solitons, with arbitrary amplitude $\eta $,
are generated by the NLSE solitons,%
\begin{equation}
U\left( z,t\right) =\eta \exp \left( i\left( \eta ^{2}/2\pm \sqrt{1-\gamma
^{2}}\right) z\right) \mathrm{sech}\left( \eta t\right) .  \label{sol}
\end{equation}

As concerns stability of the solitons in this system, it is relevant to
compare it to the stability in the usual coupler model, with $\gamma =0$. As
explained above (see Eq. (\ref{E2})), the symmetric solitons in the
nonlinear coupler are unstable against the spontaneous symmetry breaking at
\begin{equation}
\eta ^{2}>\eta _{\max }^{2}\left( \gamma =0\right) \equiv 4/3  \label{4/3}
\end{equation}
\cite{Wabnitz}, while antisymmetric solitons are always unstable \cite{Akhm2}
(although their instability may be weak).

The analysis of the instability of the usual two-component symmetric
solitons against antisymmetric perturbations, $\delta u=-\delta v$, which
leads to the exact result (\ref{4/3}), see Eqs. (\ref{anti}) and (\ref{U1})
above, can be extended for the $\mathcal{PT}$-symmetric system. The
respective perturbation $\delta u$ at the critical point, $\eta ^{2}=\eta
_{\max }^{2}$, obeys the linearized equation,
\begin{equation}
\left\{ 4\sqrt{1-\gamma ^{2}}-d^{2}/dt^{2}+\eta _{\max }^{2}\left[ 1-6%
\mathrm{sech}^{2}\left( \eta _{\max }t\right) \right] \right\} \delta u=0,
\label{U1PT}
\end{equation}%
which is the respective generalization of Eq. (\ref{U1}). This equation with
the P\"{o}schl-Teller potential is solvable, yielding
\begin{equation}
\eta _{\max }^{2}\left( \gamma \right) =(4/3)\sqrt{1-\gamma ^{2}}.
\label{crit}
\end{equation}

This analytical prediction was verified by direct simulations of the
perturbed evolution of the $\mathcal{PT}$-symmetric solitons. The
simulations were run by adding finite initial antisymmetric perturbations,
at the amplitude level of $\pm 3\%$, to the symmetric solitons. For $%
\mathcal{PT}$-antisymmetric solitons, the instability boundary was
identified solely in the numerical form, by running the simulations with
initial symmetric perturbations. The results are summarized in Fig. \ref%
{fig15}, as per Ref. \cite{PT1}. The numerically identified stability border
for the symmetric solitons goes somewhat below the analytical one (\ref{crit}%
) because the finite perturbations used in the simulations are actually not
quite small. Taking smaller perturbations, one can obtain the numerical
stability border approaching the analytical limit. For instance, at $\gamma
=0.5$, the perturbations with relative amplitudes $\pm 5\%$, $\pm 3\%$, and $%
\pm 1\%$ give rise to the stability border at $\eta _{\max }^{2}=1.02$, $%
1.055$, and $1.08$, respectively, while Eq. (\ref{crit}) yields $\eta _{\max
}^{2}\approx 1.15$ in the same case. As concerns the $\mathcal{PT}$%
-antisymmetric solitons, a detailed analysis demonstrated that they are
completely unstable \cite{PT2}, while the numerically found boundary
delineates an area in which the instability is very weak.

\begin{figure}[tbh]
\centerline{\includegraphics[width=8.0cm]{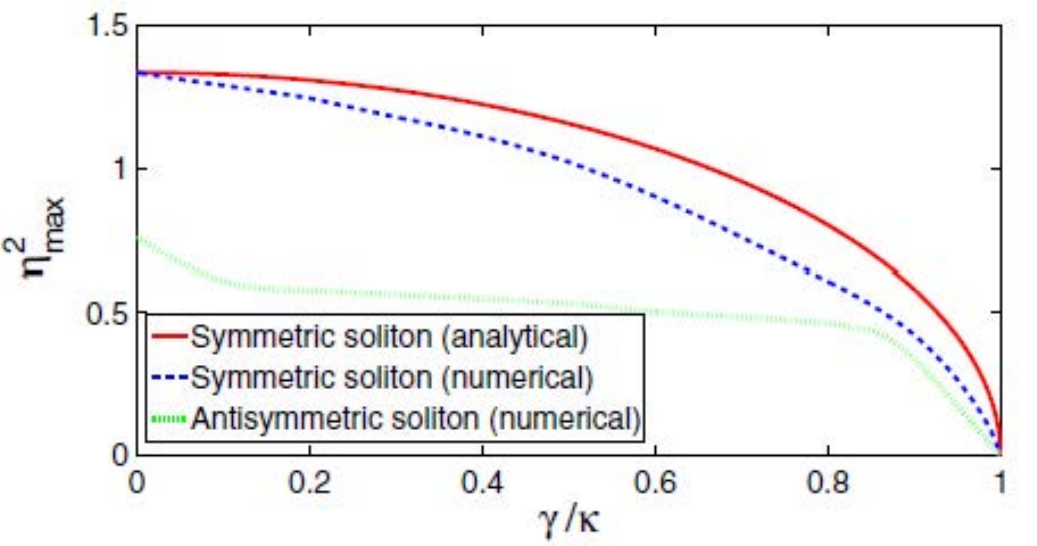}}
\caption{The analytically predicted stability border (\protect\ref{crit})
for the $\mathcal{PT}$-symmetric solitons, and its counterpart produced by
systematic simulations of the perturbed evolution of the solitons, as per
Ref. \protect\cite{PT1}. An effective numerical stability border for the $%
\mathcal{PT}$-antisymmetric solitons is shown too, although all the
antisymmetric solitons are, strictly speaking, unstable \protect\cite{PT2}
(the numerically identified stability area for them implies very weak
instability). The solitons with amplitude $\protect\eta $ are stable at $%
\protect\eta <\protect\eta _{\max }$. In this figure, $\protect\gamma /%
\protect\kappa $ is identical to $\protect\gamma $, as the inter-core
coupling coefficient is fixed by scaling to be $\protect\kappa =1$, see the
text.}
\label{fig15}
\end{figure}

It is relevant to stress that, being the stability boundary of the $\mathcal{%
PT}$-symmetric solitons, the present system, unlike the usual dual-core
coupler (see above), cannot support asymmetric solitons, as the balance
between the gain and loss is obviously impossible for them. Accordingly, the
instability of solitons with $\eta >\eta _{\max }$ leads to a blowup of the
pumped field, $u$, and decay of the attenuated one, $v$, in direct
simulations (not shown here).

\begin{figure}[tbh]
\centerline{\includegraphics[width=9.5cm]{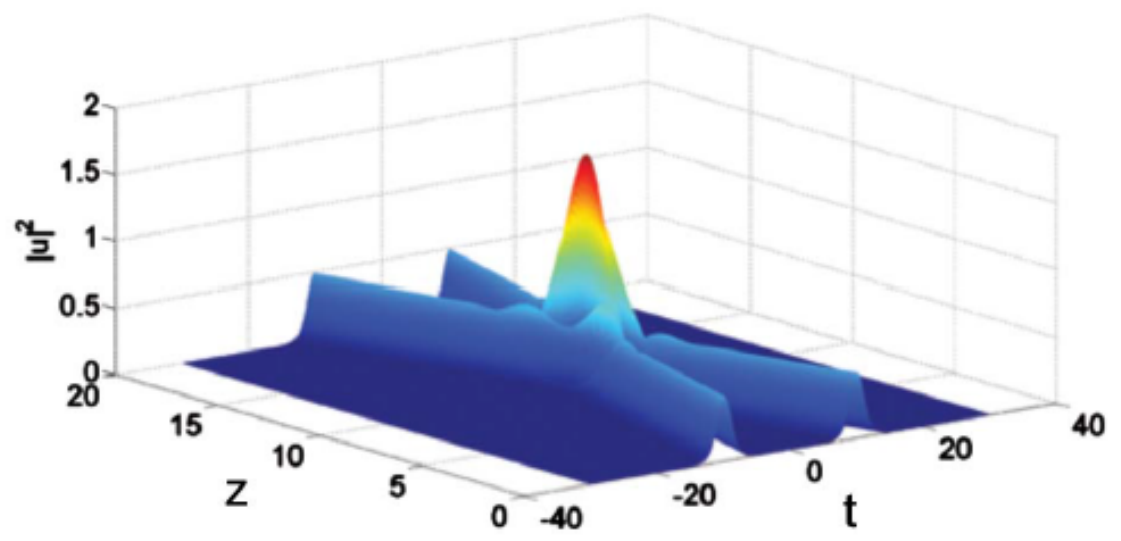}}
\caption{The elastic collision between stable $\mathcal{PT}$-symmetric
solitons with $\protect\eta =0.7$, boosted by frequency shift $\protect\chi %
=\pm 5$ at $\protect\gamma =0.7$, see Eq. (\protect\ref{boost}). The figure
is shown as per Ref. \protect\cite{PT1}.}
\label{fig16}
\end{figure}

The presence of the gain and loss terms in Eqs. (\ref{uPT}) and (\ref{vPT})
does not break their Galilean invariance, which suggests to consider
collisions between moving stable solitons, setting them in motion by means
of \emph{boosting}, i.e., replacing
\begin{equation}
\left\{ u,v\right\} \rightarrow \left\{ u,v\right\} \exp (\pm i\chi t)
\label{boost}
\end{equation}
in the initial state ( $z=0$), with arbitrary frequency shift $\chi $.
Simulations demonstrate that the collisions are always elastic, see a
typical example in Fig. \ref{fig16}.

The limit case of $\gamma =1$ may be considered as one of ``supersymmetry",
because the inter-core coupling constant and gain-loss coefficients are
equal in Eqs. (\ref{uPT}) and (\ref{vPT}) in this case. According to Eq. (%
\ref{crit}), the stability region of the solitons in the supersymmetric
systems shrinks to nil. The linearization of Eqs. (\ref{uPT}) and (\ref{vPT}%
) with $\gamma=1$ around the NLSE solution (\ref{UU}) leads to the following
equations for perturbations $\delta u$ and $\delta v$:
\begin{gather}
\hat{L}\left( \delta u+i\delta v\right) =0,  \label{LLL1} \\
\hat{L}\left( \delta u-i\delta v\right) =-2i\kappa \left( \delta u+i\delta
v\right) ,  \label{LLL2}
\end{gather}%
where $\hat{L}\delta u\equiv \left[ i\partial _{z}+(1/2)\partial
_{tt}+2|U|^{2}\right] \delta u+U^{2}\delta u^{\ast }$ is the NLSE\
linearization operator. If the underlying NLSE solution is stable by itself,
Eq. (\ref{LLL1}) produces no instability, while Eq. (\ref{LLL2}) gives rise
to a \textit{resonance}, as $\left( \delta u+i\delta v\right) $ is a zero
mode of operator $\hat{L}$. According to the linear-resonance theory \cite%
{LL2}, the respective perturbation $\left( \delta u-i\delta v\right) $ is
unstable, growing $\sim z$ (rather than exponentially). Direct simulations
of Eqs. (\ref{uPT}) and (\ref{vPT}) with $\gamma =1$ confirm that the
solitons are unstable, the character of the instability being consistent
with its subexponential character \cite{PT1}.

The ``supersymmetric" solitons may be stabilized by means of the \emph{%
management} technique \cite{book}, which, in the present case, periodically
reverses the common sign of the gain-loss and inter-core coupling
coefficients, between $\gamma =K=+1$ and $-1$ (recall $K$ is the coefficient
of the inter-core coupling, which was scaled to be $1$ above, and may now
jump between $+1$ and $-1$). Flipping $\gamma $ between $+1$ and $-1$
implies switch of the gain between the two cores, which is possible in the
experiment. The coupling coefficient, $\kappa $, cannot flip by itself, but
the signal in one core may pass $\pi $-shifting plates, which is tantamount
to the periodic sign reversal of $K$.

Ansatz (\ref{UU}) still yields an exact solutions of Eqs. (\ref{uPT}) and (%
\ref{vPT}) with coefficients $\gamma =K$ subjected to the periodic
management. On the other hand, the replacement of $K$ by the periodically
flipping coefficient destroys the resonance in Eq. (\ref{LLL2}). In
simulations, this management mode indeed maintains robust supersymmetric
solitons \cite{Radik}.

\section{Conclusion}

Dual-core optical fibers is a research area which gives rise to a great
variety of topics for fundamental theoretical and experimental studies, as
well as to a plenty of really existing and potentially possible applications
to photonics, including both traditional optics and plasmonics. While
currently employed devices based on dual-core waveguides operate in the
linear regime (couplers, splitters, etc.), the use of the intrinsic
nonlinearity offers many more options, chiefly related to the use of
self-trapped robust modes in the form of solitons. In terms of fundamental
studies, solitons in dual-core fibers are the subject of dominant interest.

Theoretical studies of solitons in these systems had begun about three
decades ago \cite%
{switch1,switch2,switch3,fiber-coupler2,Trillo,Wabnitz,Laval,Snyder,
Maimistov,Herb1,Herb2,Wabnitz2,Akhm,Akhm2,UNSW,Ank,Akhmed}. The earlier
works and more recent ones have produced a great advancement in this field,
with the help of analytical and numerical methods alike. The present article
is focused on reviewing the results obtained, chiefly, in three most
essential directions: the spontaneous symmetry breaking of solitons in
couplers with identical cores and the formation of asymmetric solitons; the
creation of stable dissipative solitons in gain-carrying nonlinear fibers
(actually, fiber lasers), stabilized by coupling the active (pumped) core to
a parallel lossy one; and the stability of solitons in $\mathcal{PT}$%
-symmetric couplers, with equal strengths of gain and loss carried by the
parallel cores.

While the analysis of these areas has been essentially completed, taking
into account both early and recent theoretical results, there remain many
directions for the extension of the studies. In particular, a natural
generalization of dual-core fibers is provided by multi-core arrays, which
allow the creation of self-trapped modes which are discrete and continuous
along the directions across the fiber array and along the fibers,
respectively. These modes include semi-discrete solitons, which may be
expected and used in many settings\ \cite{Tur1}-\cite{RBlit}. Another
generalization implies the transition from 1D to 2D couplers, represented by
dual-core planar optical waveguides. The consideration of spatiotemporal
propagation in such a system makes it possible to predict the existence of
novel species of 2D stable ``light bullets" (spatiotemporal solitons \cite%
{bullet}), such as spatiotemporal vortices \cite{Dror}, solitons realizing
the optical emulation of spin-orbit coupling \cite{SOC,SOC-PT}, and 2D $%
\mathcal{PT}$-symmetric solitons \cite{Burlak,SOC-PT}.

The most challenging problem is that, as yet, there are very few
experimental results reported for solitons in dual-core and multi-core
systems. One of experimental findings is the creation of semi-discrete
\textquotedblleft light bullets" \cite{Jena1}, including ones with embedded
vorticity \cite{Jena2} (actually, in a transient form), in three-dimensional
arrays of fiber-like waveguides permanently written in bulk samples of
silica. Further development of experimental studies in this vast area is a
highly relevant objective.

\section*{Acknowledgment}

I thank Professor Gang-Ding Peng for his invitation to join the production
of this volume, and to write the present article.

\end{document}